\RequirePackage{fix-cm}
\documentclass[smallextended]{svjour3}       
\smartqed  

\usepackage{amsmath}
\usepackage{graphicx}
\usepackage{array}
\usepackage{caption}
\usepackage{epstopdf}
\usepackage{algorithm}
\usepackage{algorithmicx}
\usepackage{algpseudocode}
\usepackage{threeparttable}
\usepackage{subfigure}
\usepackage{multirow}
\usepackage[table,xcdraw]{xcolor}
\usepackage{diagbox}
\usepackage{multirow}
\usepackage{lineno,hyperref}
\usepackage{graphicx}
\usepackage{threeparttable}
\usepackage{booktabs}
\usepackage{hyperref}
\usepackage{url}
\usepackage{natbib}
\usepackage{graphicx}
\usepackage{amsmath}
\usepackage{enumerate}
\usepackage{multirow}
\usepackage{epstopdf}
\usepackage{array}
\usepackage{CJK}
\usepackage{float}
\usepackage{subfigure}
\usepackage{algorithmicx}
\newcommand{\PreserveBackslash}[1]{\let\temp=\\#1\let\\=\temp} \newcolumntype{C}[1]{>{\PreserveBackslash\centering}p{#1}} \newcolumntype{R}[1]{>{\PreserveBackslash\raggedleft}p{#1}} \newcolumntype{L}[1]{>{\PreserveBackslash\raggedright}p{#1}}

\usepackage{makeidx}
\usepackage[table,xcdraw]{xcolor}
\usepackage{lineno,hyperref}
\modulolinenumbers[5]
 \journalname{Scientometrics}
\begin{document}

\title{Understanding the Advisor-advisee Relationship via Scholarly Data Analysis}


\author{Jiaying Liu     \and
        Tao Tang \and
        Xiangjie Kong \and
        Amr Tolba\and
        Zafer AL-Makhadmeh\and
        Feng Xia 
}


\institute{
Corresponding author: Xiangjie Kong; \email{xjkong@ieee.org} \\
\\
        J. Liu \and
        F. Xia \and
        X. Kong\at
         Key Laboratory for Ubiquitous Network and Service Software of Liaoning Province, School of Software, Dalian University of Technology, China
          \and
           T. Tang \at
         Chengdu College, University of Electronic Science and Technology of China, China
          \and
          A. Tolba \and
          Z. AL-Makhadmeh\at
          Computer Science Department, Community College, King Saud University, Saudi Arabia
          \and
          A. Tolba\at
          Mathematics Department, Faculty of Science, Menoufia University, Egypt
}

\date{Received: date / Accepted: date}

\maketitle

\begin{abstract}
Advisor-advisee relationship is important in academic networks due to its universality and necessity. Despite the increasing desire to analyze the career of newcomers, however, the outcomes of different collaboration patterns between advisors and advisees remain unknown. The purpose of this paper is to find out the correlation between advisors' academic characteristics and advisees' academic performance in Computer Science. Employing both quantitative and qualitative analysis, we find that with the increase of advisors' academic age, advisees' performance experiences an initial growth, follows a sustaining stage, and finally ends up with a declining trend. We also discover the phenomenon that accomplished advisors can bring up skilled advisees. We explore the conclusion from two aspects: (1) Advisees mentored by advisors with high academic level have better academic performance than the rest; (2) Advisors with high academic level can raise their advisees' h-index ranking. This work provides new insights on promoting our understanding of the relationship between advisors' academic characteristics and advisees' performance, as well as on advisor choosing.
\end{abstract}

\section{Introduction}
\label{intro}
Scholarly data analytics has become a vital part of the scientific research~\citep{letchford2015advantage,lehmann2006measures,glanzel2017lexical}. Relationship extraction is a major area of interest in the scholarly social network analysis. The structure of the network usually contains a large number of nodes, with different relationships such as friendship, kinship, and hostility. Analyzing such relations can strengthen our understanding of the evolution and development of the scholarly society~\citep{xia2017big}. One of the main relations is the mentorship because mentors are important for their prot{\'e}g{\'e}s. Researchers have studied various aspects of mentorship including types, phases, outcomes and mobilities as early as 30 years ago~\citep{chao1992formal,kram1983phases}. Although there is little doubt that the study of academic mentorship is necessary, it has been neglected because there is no complete advisor-advisee dataset. Each scholar needs guidance from the advisor when she/he enters academia as a newcomer. It is widely accepted that different advisors influence advisees differently~\citep{murphy2015great}. However, where these differences are originated and how significant they are, both are worth exploring. The lack of quantitative academic mentorship analysis leads to the striking lack of specific guiding significance for advisors choosing.

The mentorship is universal and complex. Traditionally, a mentorship is defined as ``A relationship between the inexperienced mentee and an experienced senior member of a particular field"~\citep{dobson2013effective}. It has different types and forms in different fields. For instance, it can be regarded as the relationship between employees and managers in public utility companies and advisor-advisee relationship in academia. There is a large volume of studies describing mentorship in various aspects. According to the formation of the relationship,~\cite{chao1992formal} first categorized mentorship into two types: formal mentorship and informal mentorship. Informal mentorship usually has neither compact structure with reasonable management, nor been recognized by the organization formally. In contrast to informal mentorship, formal mentorship occurs with external intervention from the organization. Usually it is led and managed by the organization. Mentorship is a temporary relationship with almost fixed duration, for example, the average length of advisor-advisee relationship in academic networks is five years~\citep{kram1983phases}. The phases of the mentorship can be defined as initiation, cultivation, separation, and redefinition on the basis of the particularly effective experience~\citep{kram1983phases}. In recent years, there has been an increasing amount of literature describing specific elements of mentorship, including behaviors of mentors and prot{\'e}g{\'e}s, mentoring functions, structure of cooperation in the mentorship~\citep{johnson2015elements,bozionelos2014mentoring,borders2012association}, and so on. These studies provide complete insights into the development and implementation of the mentorship.

 No matter what the form is, the ultimate goal of mentorship is to enhance the personal and professional development of prot{\'e}g{\'e}s~\citep{chao1997mentoring}. Therefore, mentors have been taken as beacon lights in the development of prot{\'e}g{\'e}s. Based on inherently dyadic attributes of the mentorship, the existing literature demonstrates that mentorship is positive and beneficial for both mentors and prot{\'e}g{\'e}s. In a positive mentorship, the prot{\'e}g{\'e}s will have better performance and more opportunities for promotion because their mentors provide them with personal and career assistance. Meanwhile, the mentors can receive fulfillment and gain satisfaction by improving prot{\'e}g{\'e}s' welfare~\citep{ghosh2013career,hu2014mentors}. Furthermore, organizations benefit as well because there is a real possibility that prot{\'e}g{\'e}s would like to be devoted to their organizations after graduation~\citep{malmgren2010role,florea2013all}. Moreover, as early as 1989,~\cite{fagenson1989mentor} suggested that mentoring can enhance work validity and the career success.~\cite{singh2009gets} perceived to the point that compared with those who do not have a mentor, individuals who have obtain mentors embrace more chances to promote in their careers, and more likely to consider their future prospects ahead of time in more positive ways.~\cite{wanberg2006mentor} proposed sometimes a mentor may change a prot{\'e}g{\'e}'s vocational tendency which can lead the prot{\'e}g{\'e} to the different lifeline.

 The recent trends in mentorship analysis have led to a proliferation of studies focusing on the association between mentorship and the career development of prot{\'e}g{\'e}s. Since~\cite{kram1983phases} presented the psychosocial functions and career development functions to examine the mentors' influence on the prot{\'e}g{\'e}s, almost every paper focusing on the outcome of mentorship includes contents related to it. The mentoring functions have received considerable critical attention. On this basis,~\cite{chao1992formal} compared the mentoring functions among formal mentorships, informal mentorships, and non-mentored counterparts.~\cite{scandura1993effects} investigated the link between mentoring functions and the career mobility outcomes in terms of promotions and salaries. In a study conducted by~\cite{singh2009gets}, it showed the correlation of rising star attributes measured at different periods, i.e., no-mentored period and a year after mentored. In the same vein,~\cite{chao1997mentoring} examined linkages among different mentorship phases, functions and outcomes of prot{\'e}g{\'e}s in order to integrate different aspects of mentoring into a more comprehensive theory. According to~\cite{young2000did}, there were specific behaviors such as trust, related to career and social support the exhibited throughout the mentoring process. Furthermore, ~\cite{tuesta2015analysis} found the evidence that there exists a positive correlation between time of advisor-advisee relationship and advisee's productivity in the area of Exact and Earth Sciences.

 Although research has been carried out on the mentors' influence, with the exception of ~\cite{tuesta2015analysis} who measured scientific productivity by the number of publications in journals, all others~\citep{chao1992formal,scandura1993effects,singh2009gets,young2000did,chao1997mentoring} remain narrowly focus on dealing only with analytic study based on the feedback from the questionnaires. There are few data-based investigations studying the outcomes of choosing different advisors in academia. This gives rise to the question of how to choose an advisor when newcomers enter academia. In response to the question ``What kind of advisors you will choose when you decide to pursue the PhD degree?", most of the answers are non-specific and all-embracing. Almost all advisees want to choose advisors who make a great and positive influence on themselves. Young advisees in academia often seek to work in collaboration with top advisors in their field in pursuit of a successful career. However, ``top advisors" in academia can be defined differently.

 In this work, we analyze the relationship between advisees' performance (i.e., productivity and impact) and advisors' academic characteristics from the perspective of Computer Science. It is different from the existing research which only employs questionnaires. The dataset of advisor-advisee relationships is extracted from Digital Bibliography \& Library Project (DBLP) dataset~\citep{ley2009dblp}. We use the improved stacked autoencoder method based on the deep learning with the highest accuracy reaching up to 94\% to get the complete dataset~\citep{wang2016mining,wang2017shifu}. Then we make a quantitative analysis of the dataset. We calculate the productivity and impact of advisees including the number of publications, citations, and h-indices in 1-12 years after first collaborating with their advisors. We separate the advisors into groups according to their academic age to observe the relation between advisees' outcomes and advisors' academic ages. We find that advisees' number of publications, citations, and h-indices follow the same trend with the increasing of advisors' academic ages, exhibiting an initial growth, remaining stationary for a duration, finally with a declining trend. Based on the observation, we examine the correlation between the ranking of advisors and their advisees' h-indices.

 Furthermore, by combining quantitative analysis with qualitative analysis, we find that advisors with high academic levels will bring up advisees with high academic performance. We conclude the findings from two main aspects: (1) Advisees mentored by the advisors with high academic level generally have better academic performance than the rest. We divide the advisors into two groups: one with a high ranking in terms of publications/citations/h-indices and the other without. By comparing the academic performance of all advisees mentored by different advisors, we observe that the phenomenon also holds for advisees with high academic level (if a scholar ranks the top 10\% in the number of publications/citations/h-indices, then she/he has the higher academic achievement). (2) Advisors with high academic level can increase the probability of their advisees ranking the top 10\% in h-index. We calculate the probability of an advisee ranking the top 10\% in different cases. These cases are based on advisors' different h-indices and academic ages. The outcomes of this quantitative research, which compare advisees' achievements coached by advisors with the different academic performance we present here, can be utilized for advisor recommendation.

\section{Methods}

\subsection{Dataset}
Our large-scale dataset of advisor-advisee relationships is generated by applying a deep learning technique, stacked autoencoder, on the DBLP dataset~\citep{wang2016mining,wang2017shifu}. The specific steps for the advisor-advisee dataset acquisition are as follows:
\begin{enumerate}
  \item Construct the collaboration ego-network for all scientists in the DBLP dataset.
  \item Clean the network: Following Sinatra's~\citep{sinatra2016quantifying} steps, we only choose the scholars who have published at least 5 papers and who have a publication career span of at least 10 years in the DBLP dataset.
  \item Extract the required features for training as unlabeled input of the model: Compute the personal properties (i.e., academic age, number of publications) and collaboration properties (i.e., collaboration times, collaboration duration, times of first-two authors, and cohesion of collaboration) for each scholar in the first 8 years of collaboration, then use these normalized features as the input of the model. The Back Propagation (BP) method is used to train the model and optimize the model. The result of identifying advisors is obtained through classifier after training.
  \item Select the advisor-advisee pair whose identification accuracy exceeds 90\% as the advisor-advisee dataset.
\end{enumerate}
The output dataset consists of 810,469 advisor-advisee pairs and more than 1,475,000 publications from 1968 to 2016, with the highest computational precision reaching 94\%. While DBLP only provides publication details for each scholar, the accomplishment in terms of citations cannot be obtained from DBLP. In order to obtain scholars' citations and h-indices, we match our dataset with AMiner-DBLP~\citep{tang2008arnetminer}. AMiner~\footnote{https://www.aminer.cn} is a website which aims to provide comprehensive search and mining services for researcher social networks. It is considered as a widely used and one of the best-curated databases for Computer Science articles~\citep{gollapalli2011ranking,moreira2011learning,tang2008topic,amjad2017standing}. Currently, the system consists of more than 6,000 conferences, 3,200,000 publications, and 700,000 researcher profiles before 2016. In this website, developers use Microsoft Graph Search API to query each AMiner paper's title and obtain candidate matching papers for each AMiner paper. For the computational accuracy, they randomly sampled 100,000 linking pairs and evaluated the matching accuracy. The number of truly matching pairs is 99,699 and the matching accuracy can achieve 99.70\%.

\begin{figure}[htbp]
\centering
\includegraphics[width=\linewidth]{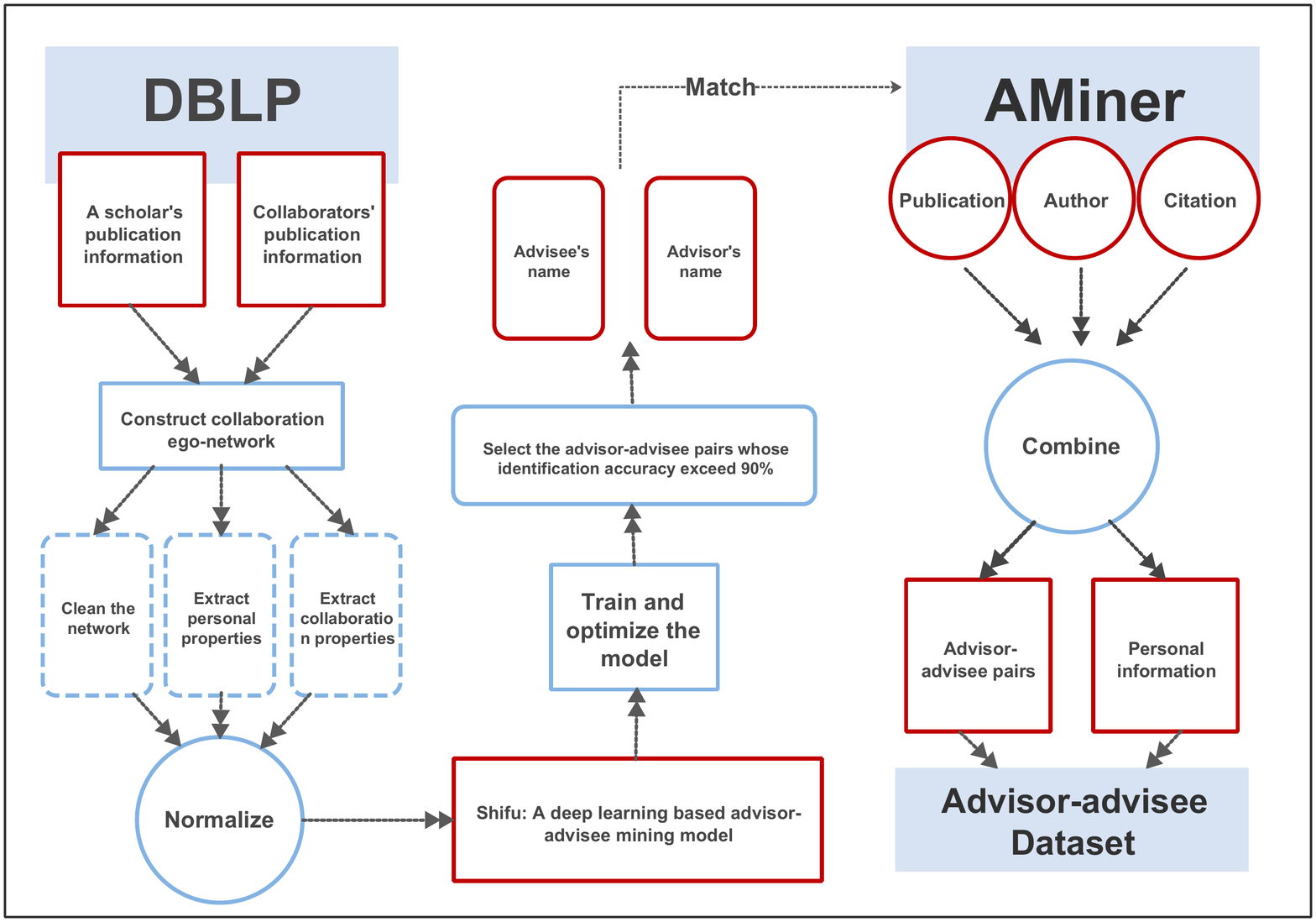}
\caption{Dataset acquisition process.}
\label{fig:11}
\end{figure}

Finally, the analysis is on the grounds of 15,559 advisors whose academic age is 5 (23,473 advisees), 20,859 advisors whose academic age is 10 (36,841 advisees), 17,028 advisors whose academic age is 15 (33,652 advisees), 11,522 advisors whose academic age is 20 (24,883 advisees), and 7,352 advisors whose academic age is 25 (11,305 advisees). The whole process of the experiment is illustrated in Fig.~\ref{fig:11}.

\subsection{Ranking the scholars}
We take three variables into consideration to rank a scholar: number of publications, citations, and h-indices. H-index of a scholar is calculated on the basis of the year of publications and referenced information each year. A scholar has an h-index $h$ if at least $h$ of his/her publications attract $h$ or more citations. In order to measure the advisors' influence on advisees during different collaboration periods, the indicators for each year are calculated.

\subsection{Analysis method}
\subsubsection{Academic age of scholars}
When dividing the type of advisors, we take advisors' academic ages into consideration because the academic characteristics such as citations and the number of publications are accumulated over time. Therefore, advisors with different academic ages should not be put together for comparison.

\begin{definition}
\label{T0.1}
AA is defined as the academic age of scholars, i.e.,
\[
\ AA = Y_c - Y_f
\]
where $Y_f$ is the year scholars published the first article and $Y_c$ is the investigated year.
\end{definition}

In this paper, we calculate advisees' and advisors' academic ages of the year in which they began cooperating with each other. In other words, $Y_c$ is the year when they commenced collaboration.

\subsubsection{The probability of advisees whose h-indices rank the top 10\%}
In order to observe the association between advisees' academic performance and advisors' academic level, we calculate the possibility that advisees mentored by different types of advisors with different h-index rankings and academic ages. The steps are described as follows:
\\(1) Divide the advisors according to the academic age when they first collaborated with their advisees. For advisors with the same academic age, divide them into $Top_{10}$ group in which advisors' h-index ranking is in the top 10\% and $Res$ group for the rest.
\\(2) Calculate the total number of $Top_{10}$ advisors' advisees $Nt_t$ and $Res$ advisors' advisees $Nt_r$.
\\(3) Rank all the advisees with the same academic age according to their h-indices.
\\(4) Calculate the minimum of the top 10\% advisees' h-index $h_{min}$.
\\(5) Calculate the number of $Top_{10}$ advisors' advisees $Nh_t$ whose h-indices are above $h_{min}$. In the same way, calculate $Nh_r$ of $Res$ advisors.
\\(6) The probability of $Top_{10}$ advisors' advisees whose h-index rank top 10\% can be calculated as $ Nh_t / Nt_t$ and $ Nh_r / Nt_r$ for $Res$ advisors.

\section{Results}
\label{sec:1}
In this section, we present the basic analysis of the advisor-advisee dataset in subsection 3.1. Subsection 3.2 elaborates the correlation between advisors' academic characteristics and advisees' academic performance. Subsection 3.3 shows the phenomenon that an accomplished advisor could bring up a skilled advisee.
Finally, the analysis of the increasing publication rate is highlighted in subsection 3.4.

\subsection{Statistical analysis of the advisor-advisee relationships}
Table \ref{tab:1} summarizes the statistics of the advisors at the beginning of collaborating with their advisees. It illustrates the mean of advisors' academic characteristics including the number of publications ($NP$), citations ($NC$), and h-indices ($NH$). Here we separate advisors according to their exact academic age ($AA$) when they begin guiding the advisees, to mitigate the bias of accumulated citations for advisors with different $AA$.

\begin{table}[htbp]
\centering
\caption{Academic characteristics of advisors}
\label{tab:1}
\begin{tabular}{l|cccccc}
\hline
Attributes& $AA=5$ & $AA=10$ & $AA=15$ & $AA=20$ & $AA=25$ & Average \\
\hline
$NP$& 12.21  & 17.65 & 12.01 & 10.22 & 9.91& 11.73\\
$NC$& 57.11  & 145.2 & 195.05& 272.74 & 319 & 160.4\\
$NH$&  3.13  & 5.29&  5.93&  6.84& 7.25&  4.88\\
\hline
\end{tabular}
\end{table}

\begin{figure}[htbp]
\centering
\subfigure[Academic ages]{
\label{fig:1-a}
\includegraphics[width=0.47\textwidth]{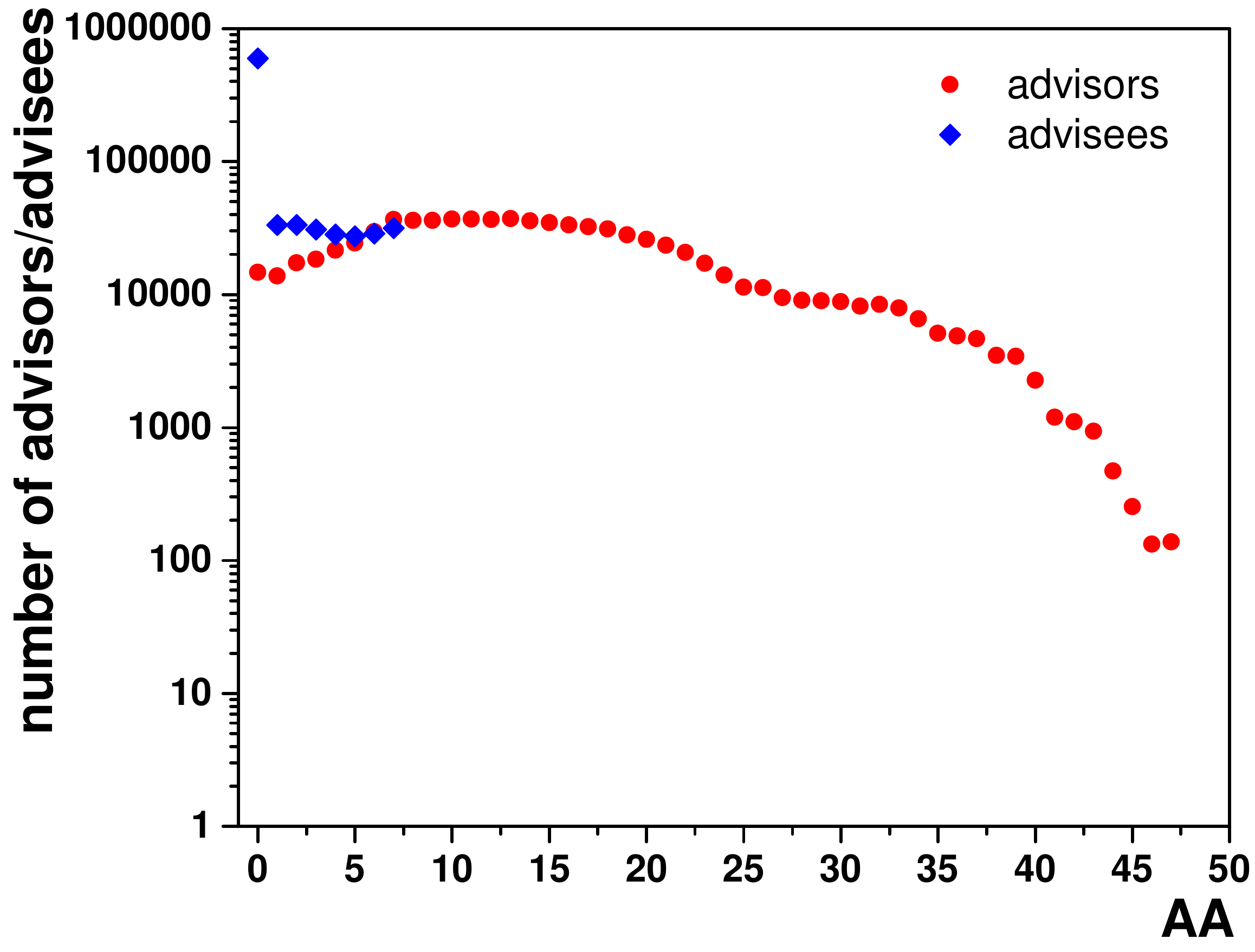}}
\subfigure[Survival function for $AA$]{
\label{fig:1-b}
\includegraphics[width=0.435\textwidth]{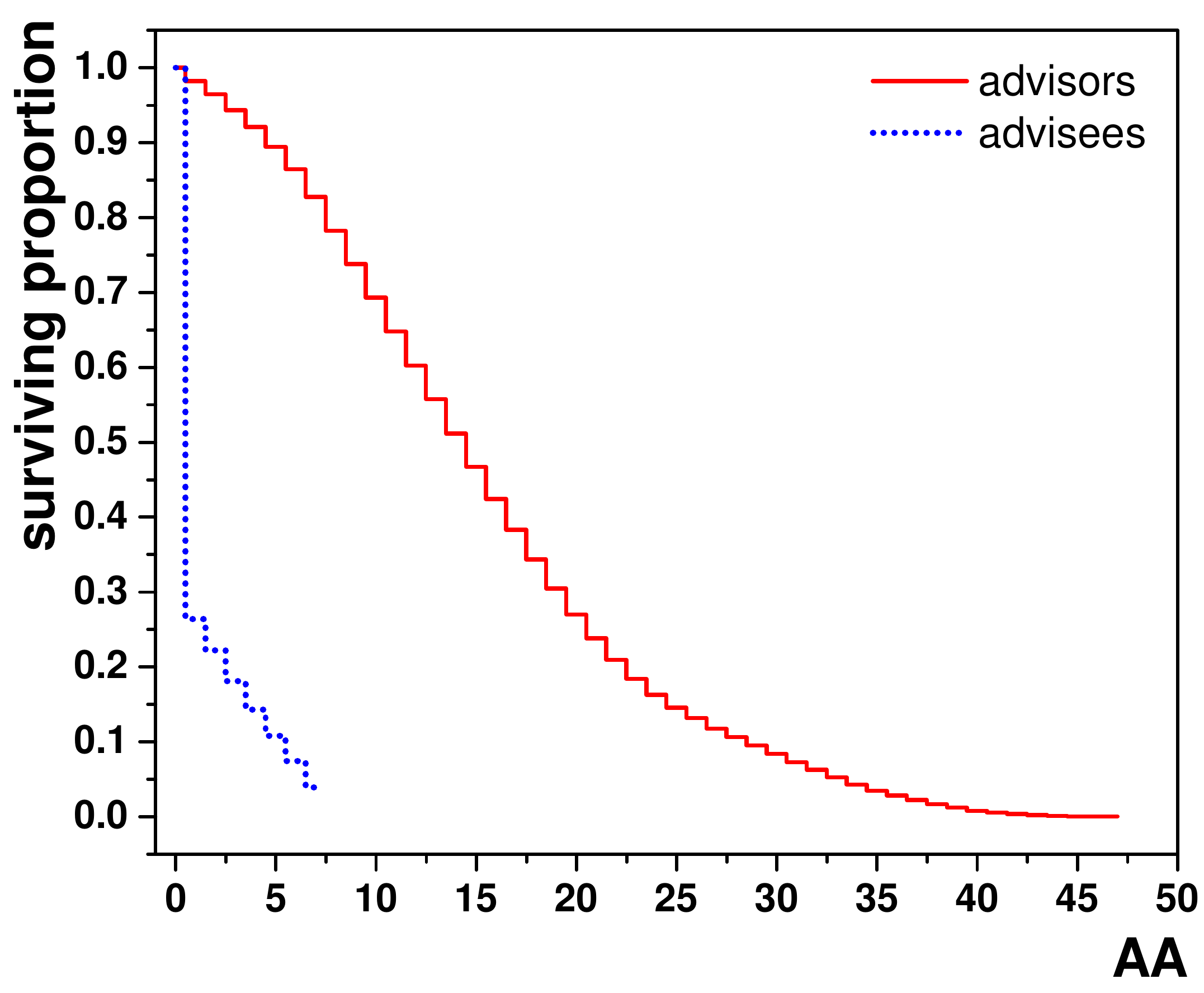}}\\
\subfigure[Number of publications]{
\label{fig:1-c}
\includegraphics[width=0.47\textwidth]{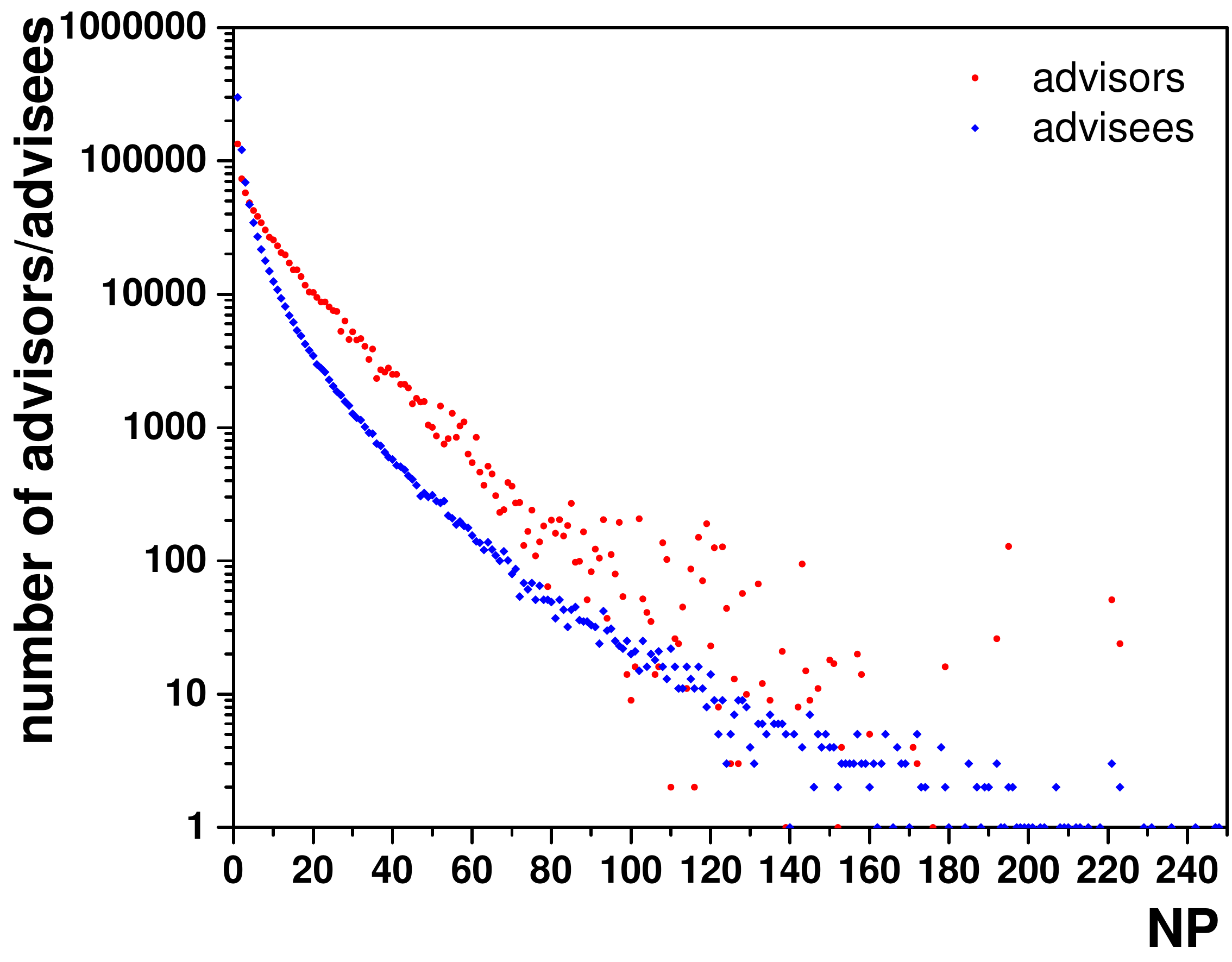}}
\subfigure[Survival function for $NP$]{
\label{fig:1-d}
\includegraphics[width=0.45\textwidth]{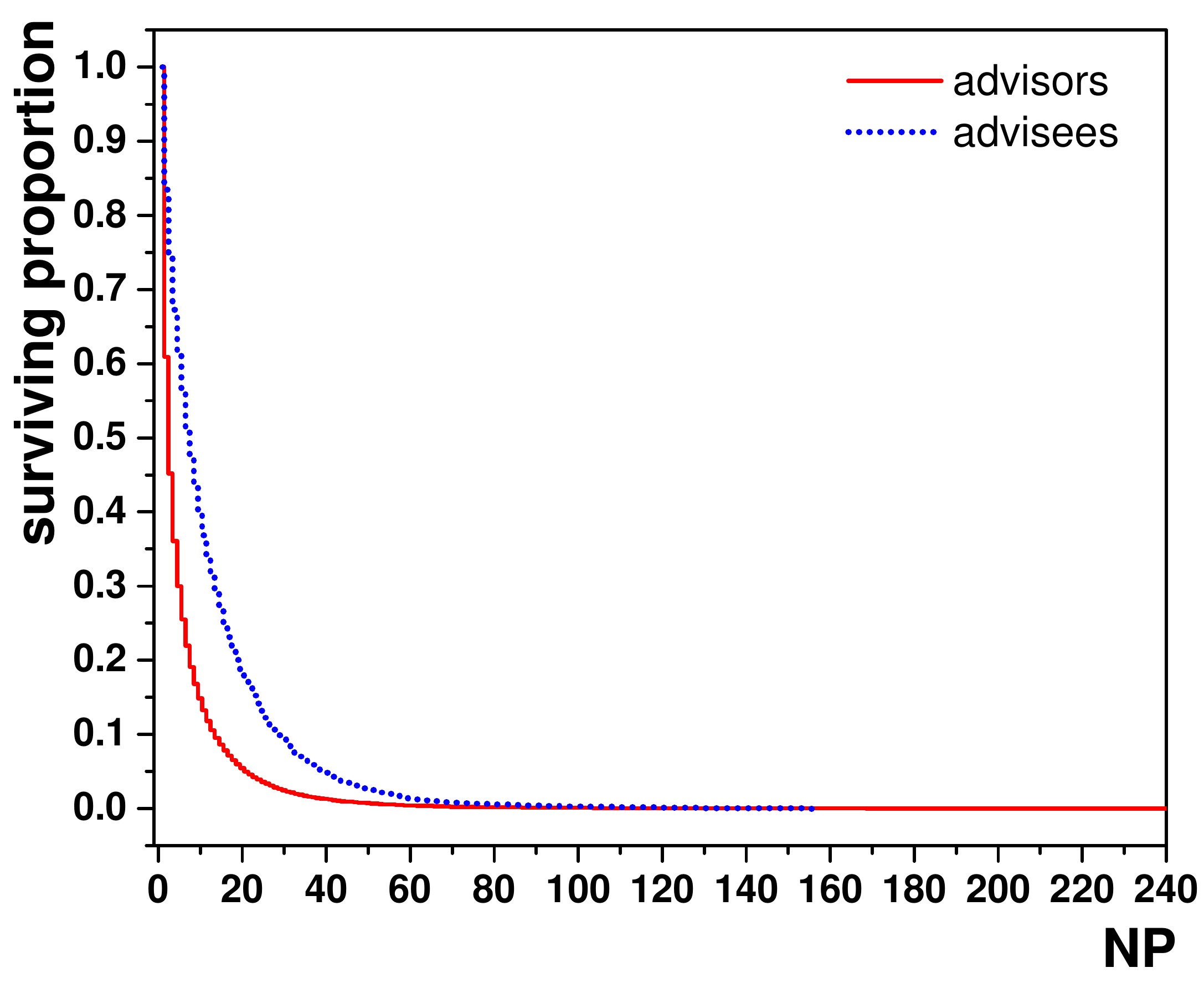}}\\
\subfigure[Citations]{
\label{fig:1-e}
\includegraphics[width=0.47\textwidth]{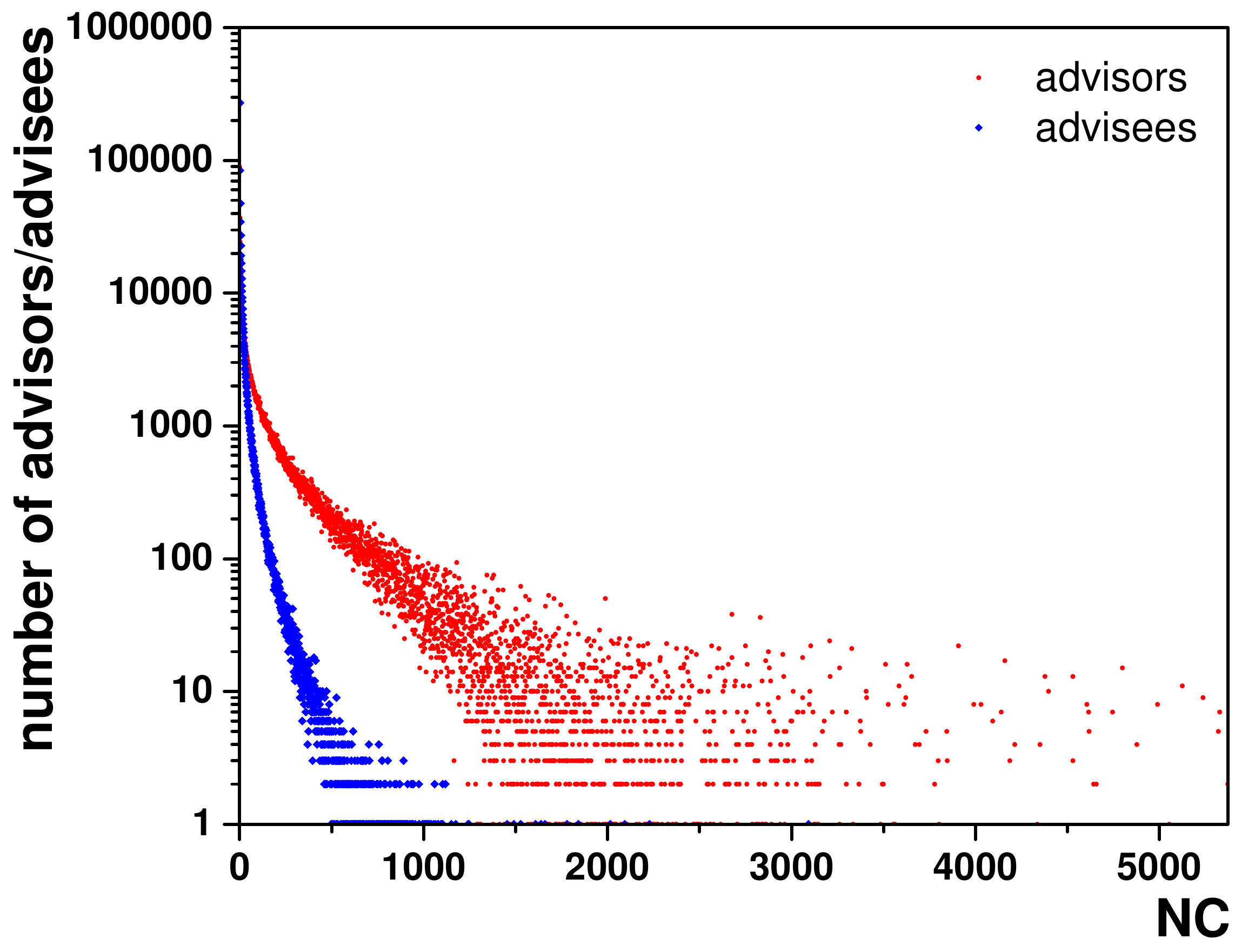}}
\subfigure[Survival function for $NC$]{
\label{fig:1-f}
\includegraphics[width=0.44\textwidth]{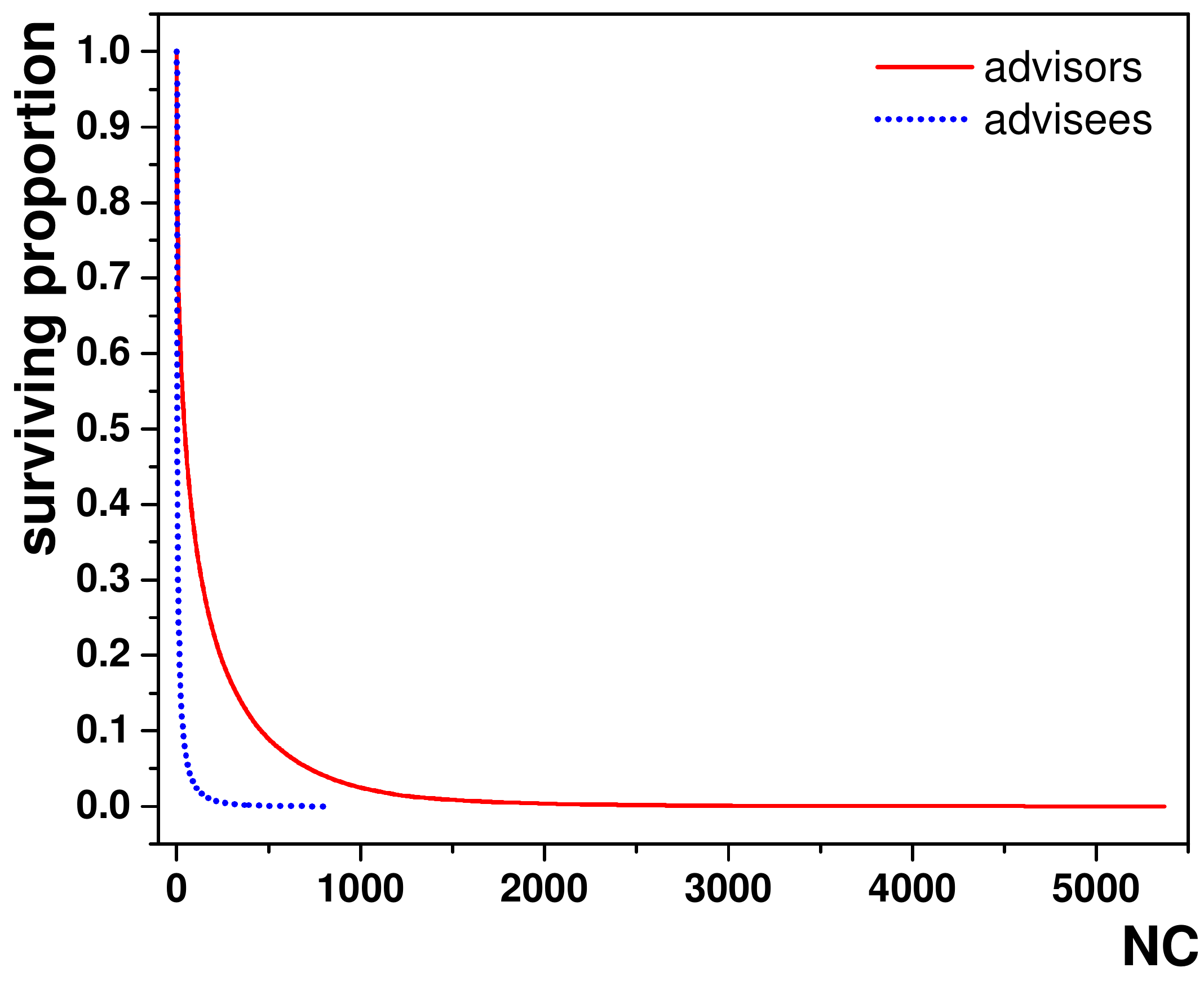}}\\
\subfigure[H-indices]{
\label{fig:1-g}
\includegraphics[width=0.46\textwidth]{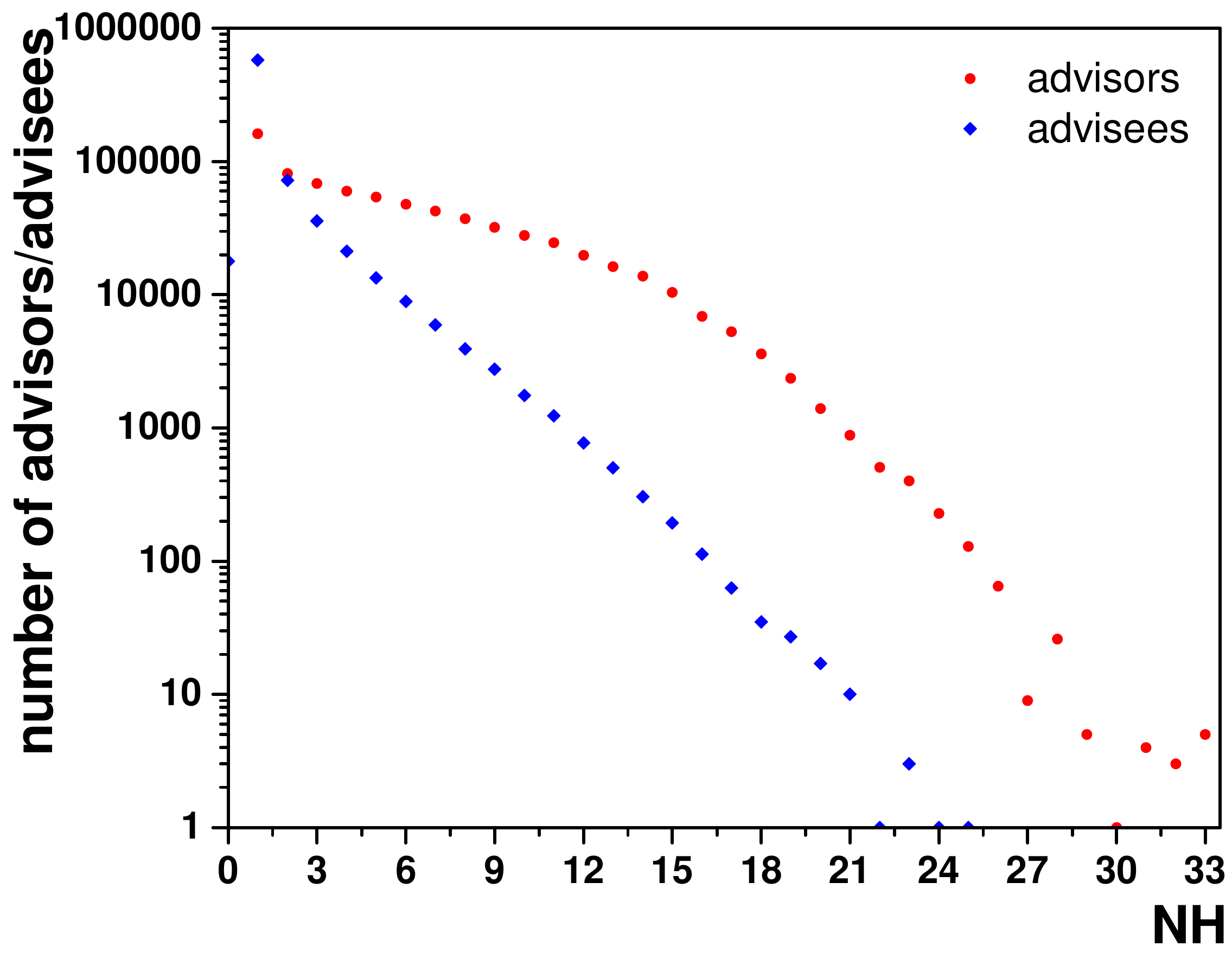}}
\subfigure[Survival function for $NH$]{
\label{fig:1-h}
\includegraphics[width=0.44\textwidth]{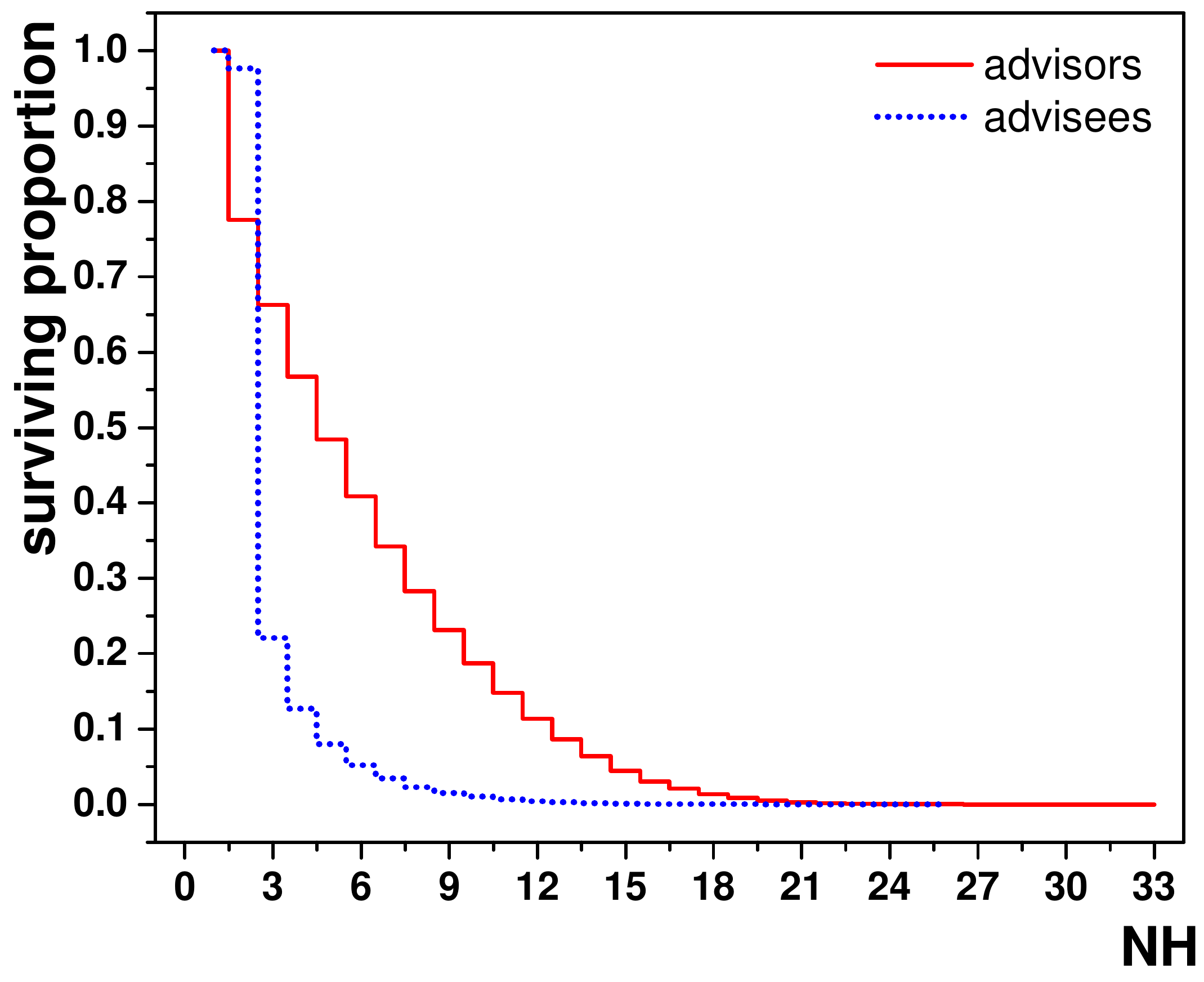}}
\caption{Academic performance distribution and survival function for advisors and advisees. The horizontal axis represents (a) academic ages, (c) the number of publications, (e) citations, and (g) h-indices, respectively. The vertical axis represents the number of scholars. In (b), (d), (f), and (h), the vertical axis represents the surviving proportion.}
\label{fig:1}
\end{figure}

In order to carry out the data-based analysis of the advisor-advisee relationships, we keep a tally of relevant information of advisors and advisees in the dataset. Fig. \ref{fig:1} shows the range of each personal characteristic and number of advisors/advisees distribution of these characteristics.

The red points represent advisors and blue ones represent advisees. Fig. \ref{fig:1-a} shows the academic age distribution of advisors and advisees since their first collaboration. As shown in Fig. \ref{fig:1-a}, most advisors' $AA$ are 5-20 years. After that, as $AA$ increases, the number of advisors decreases. Fig. \ref{fig:1-c}, Fig. \ref{fig:1-e}, and Fig. \ref{fig:1-g}, respectively illustrate the distributions of advisees (10 years after first collaborating with their advisors) and advisors (when they began mentoring a certain advisee) in terms of $NP$, $NC$, and $NH$. We only consider the advisees whose career spans at least 10 years since they first collaborated with their advisors. It can be found from Fig.~\ref{fig:1-c}, Fig. \ref{fig:1-e}, and Fig. \ref{fig:1-g}, that there is a clear decreasing trend in the number of scholars with the increasing of $NP$, $NC$, and $NH$. We also plot the survival function, that is $P (x > X)$ for these graphs, which will get rid of the noise and observe the differences between various groups. The results are presented in Fig. \ref{fig:1-b}, Fig. \ref{fig:1-d}, Fig. \ref{fig:1-f}, and Fig. \ref{fig:1-h}. In these figures, the horizontal axis represents scholars' $AA$, $NP$, $NC$, and $NH$, respectively. The vertical axis is the survival rate. These survival curves have a declining trend. The steeper decline slope is on behalf of the lower survival rate. From these figures, we can see that all curves are concave. It illustrates that most scholars have relatively lower academic performance, few scholars have higher academic achievements, which can be evidenced by ``Long Tail Effect"~\citep{anderson2006long}.

\begin{table}[htbp]
\centering
\caption{Exponential fitting of $AA$, $NP$, $NC$, and $NH$ distribution for advisors and advisees}
\label{tab:2}
\begin{tabular}{cc}
\hline
Attributes & Fitting Function  \\
\hline
Advisees' $AA$ & $ y = y_{0} + Ae^{R_{0}x}$ \\
Advisors' $AA$ & $ y = Ae^{-AX}$ \\
Advisees' $NP$ & $ y = ae^{bx}$ \\
Advisors' $NP$ & $ y = y_{0} + A_{1}e^{\frac{x}{t_{1}}}$ \\
Advisees' $NC$ & $ y = y_{0} + \frac{A}{w\sqrt{\frac{\pi}{2}}}e^{-2\frac{(x-x_{c})^{2}}{w^{2}}}$ \\
Advisors' $NC$ & $ y = y_{0} + A_{1}e^{\frac{x}{t_{1}}}+ A_{2}e^{\frac{x}{t_{2}}} + A_{3}e^{\frac{x}{t_{3}}}$  \\
Advisees' $NH$ & $ y = y_{0} + \frac{A}{w\sqrt{\frac{\pi}{2}}}e^{-2\frac{(x-x_{c})^{2}}{w^{2}}}$  \\
Advisors' $NH$ & $ y = y_{0}+ A_{1}e^{\frac{x}{t_{1}}}$  \\
\hline
\end{tabular}
\end{table}

\begin{figure}[htbp]
\centering
\subfigure[$AA$ of advisees]{
\label{fig:2-a}
\includegraphics[width=0.47\textwidth]{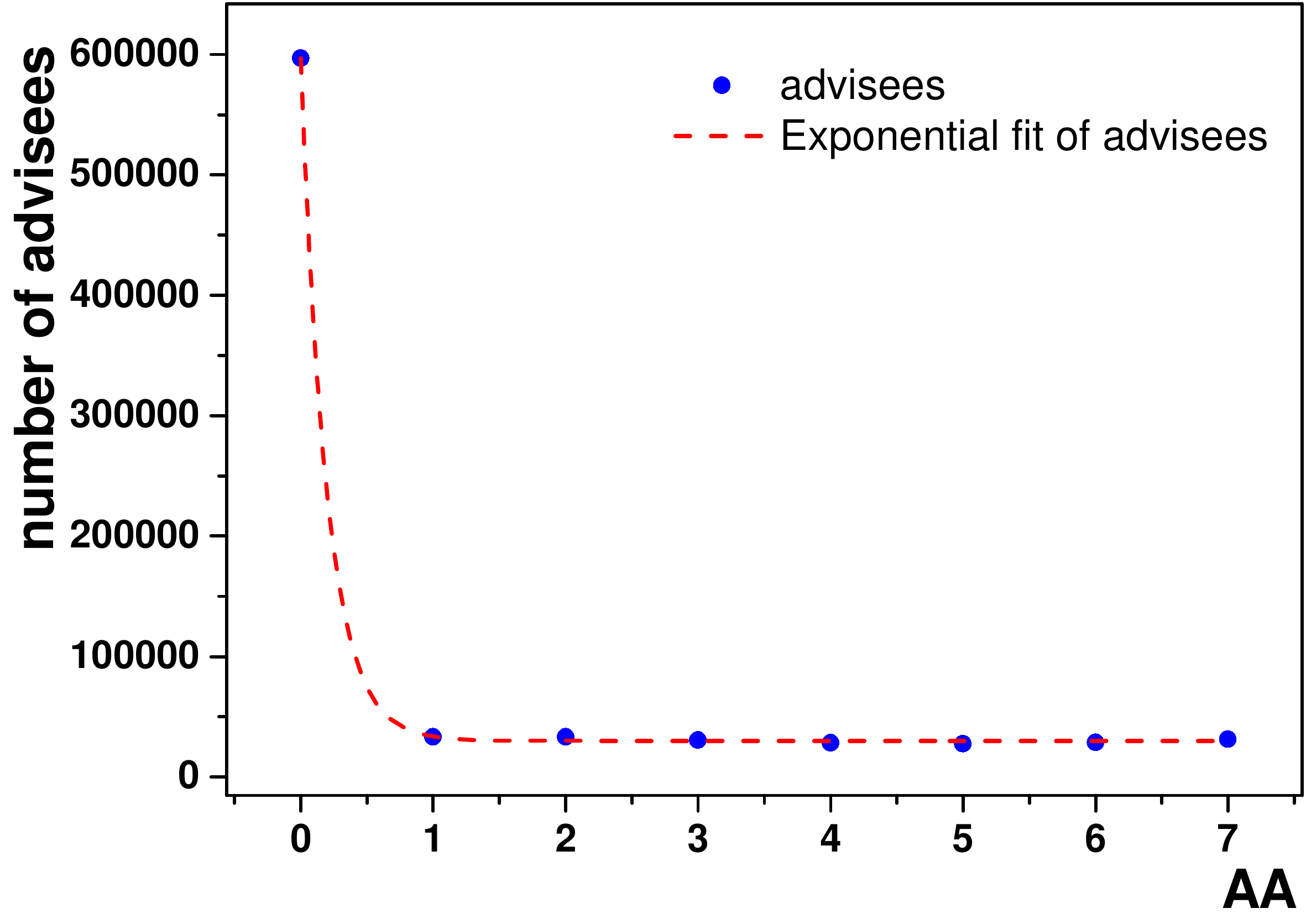}}
\subfigure[$AA$ of advisors]{
\label{fig:2-b}
\includegraphics[width=0.46\textwidth]{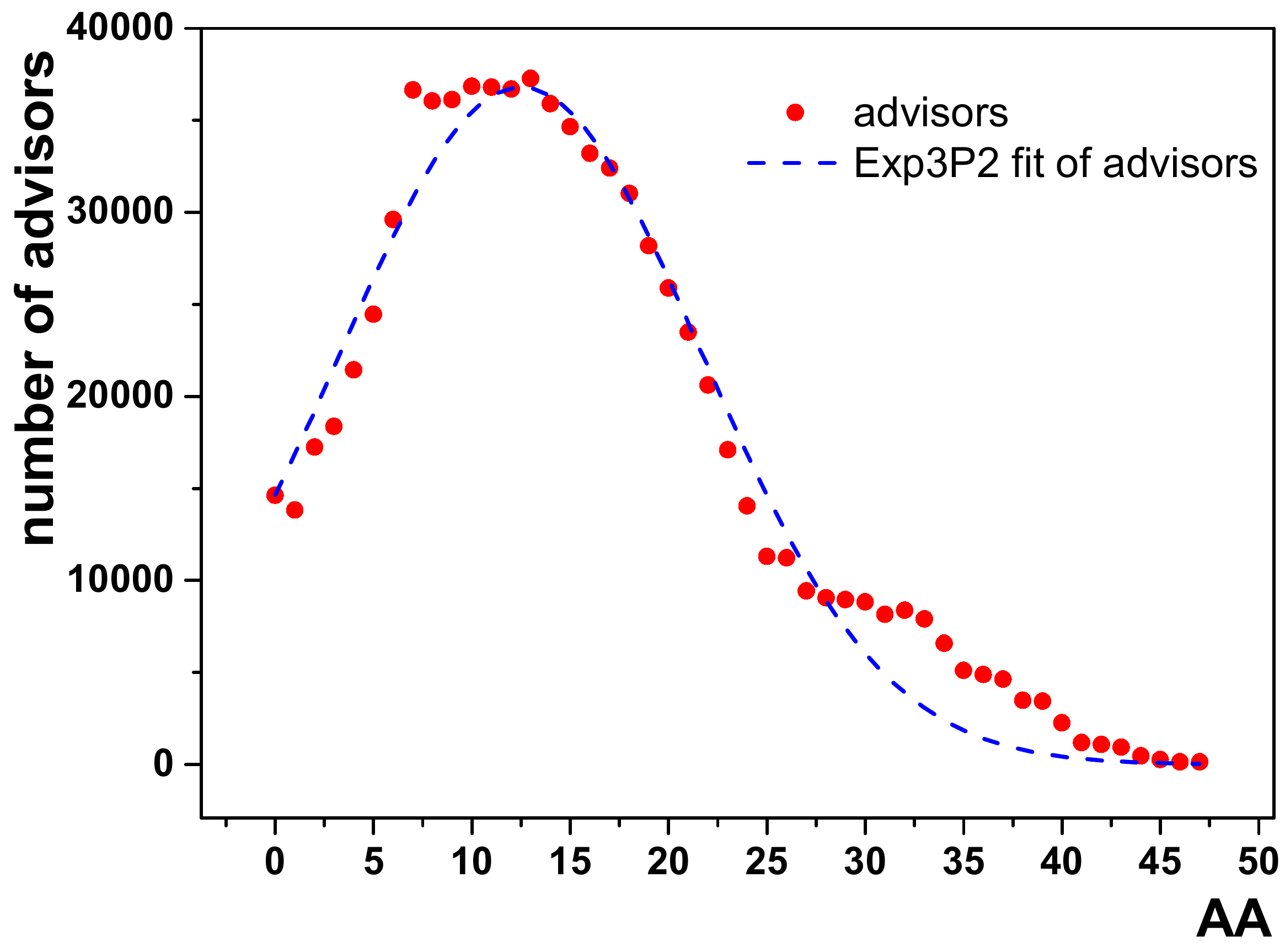}}\\
\subfigure[$NP$ of advisees]{
\label{fig:2-c}
\includegraphics[width=0.47\textwidth]{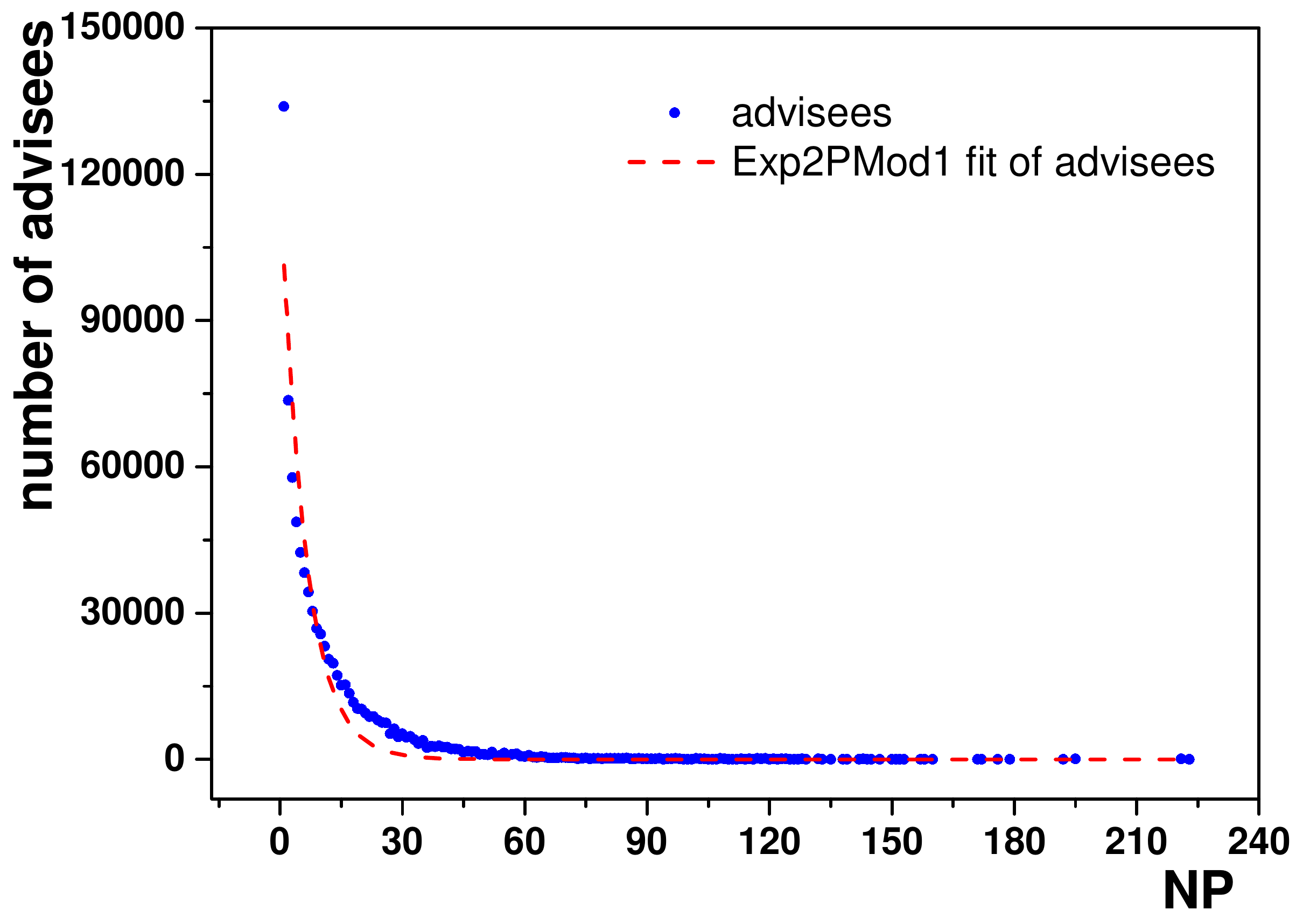}}
\subfigure[$NP$ of advisors]{
\label{fig:2-d}
\includegraphics[width=0.47\textwidth]{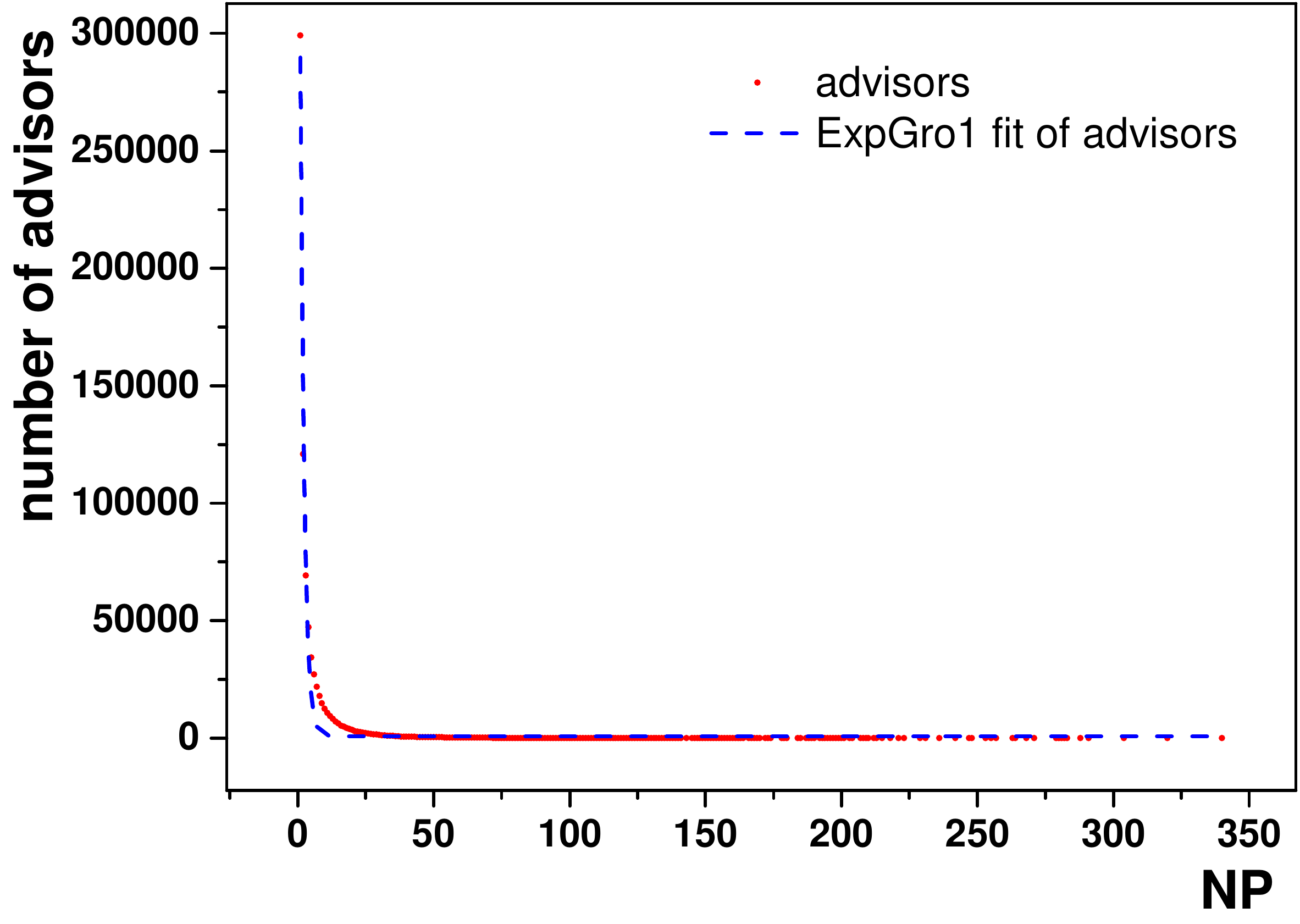}}\\
\subfigure[$NC$ of advisees]{
\label{fig:2-e}
\includegraphics[width=0.47\textwidth]{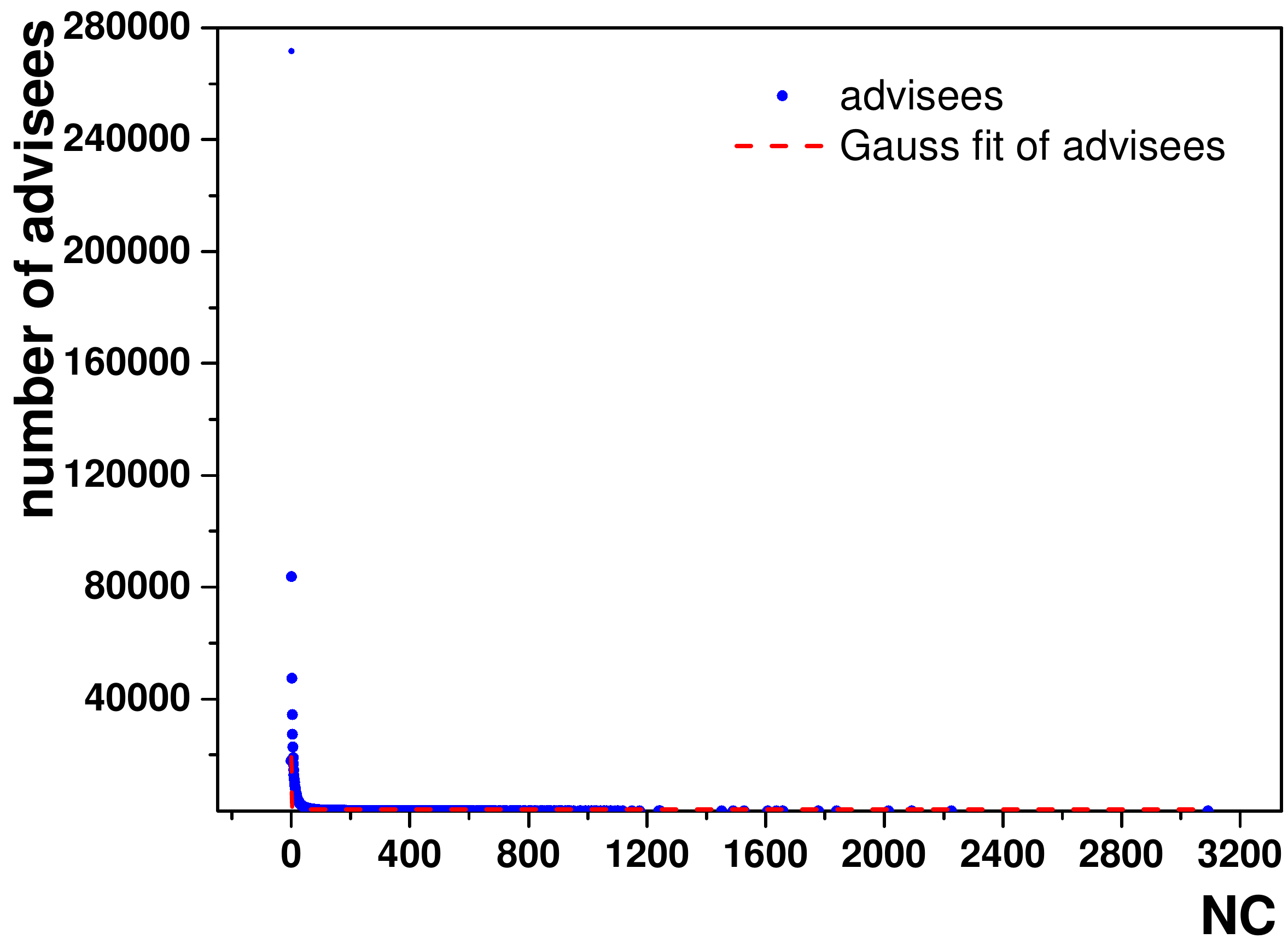}}
\subfigure[$NC$ of advisors]{
\label{fig:2-f}
\includegraphics[width=0.47\textwidth]{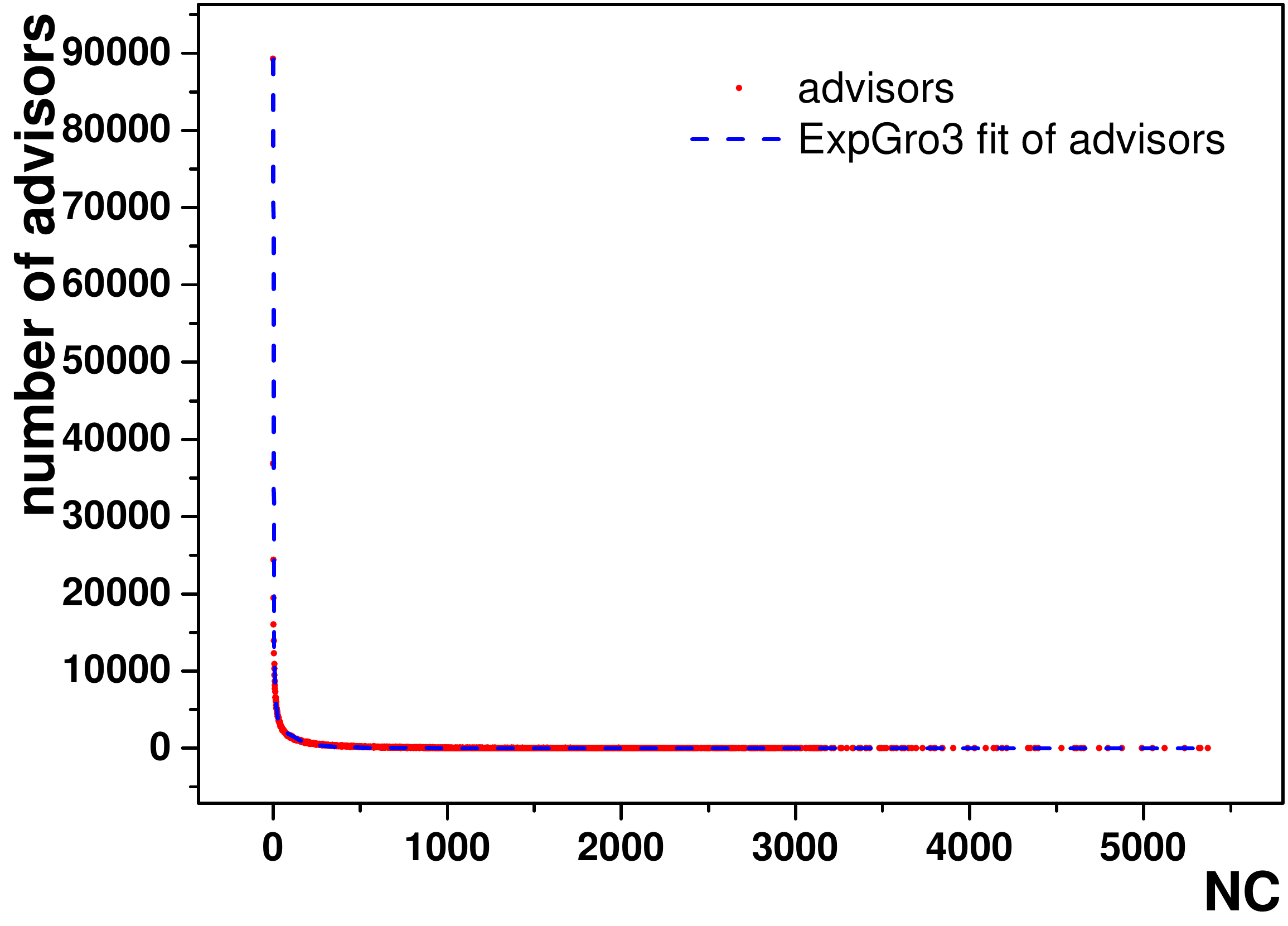}}\\
\subfigure[$NH$ of advisees]{
\label{fig:2-g}
\includegraphics[width=0.47\textwidth]{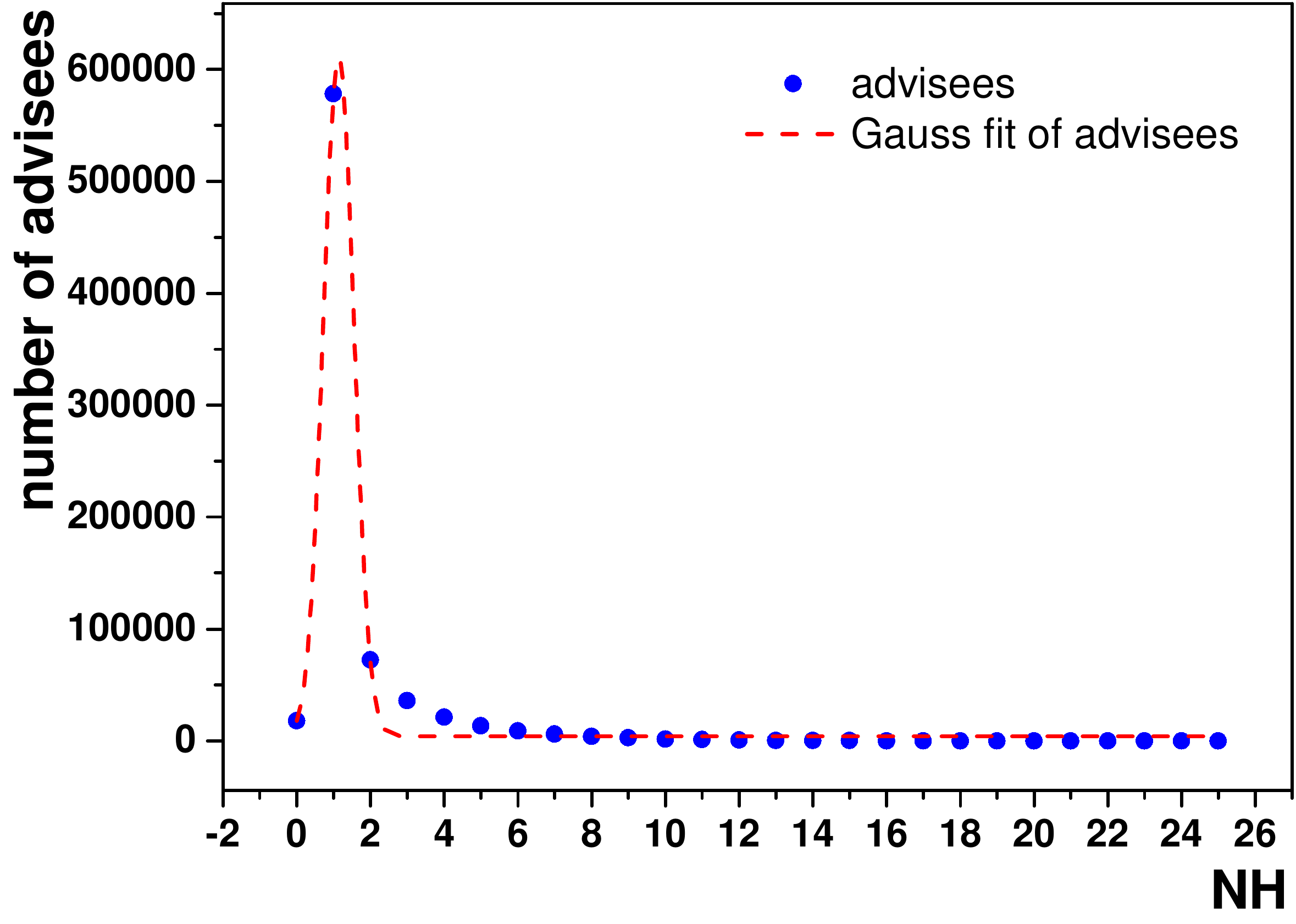}}
\subfigure[$NH$ of advisors]{
\label{fig:2-h}
\includegraphics[width=0.49\textwidth]{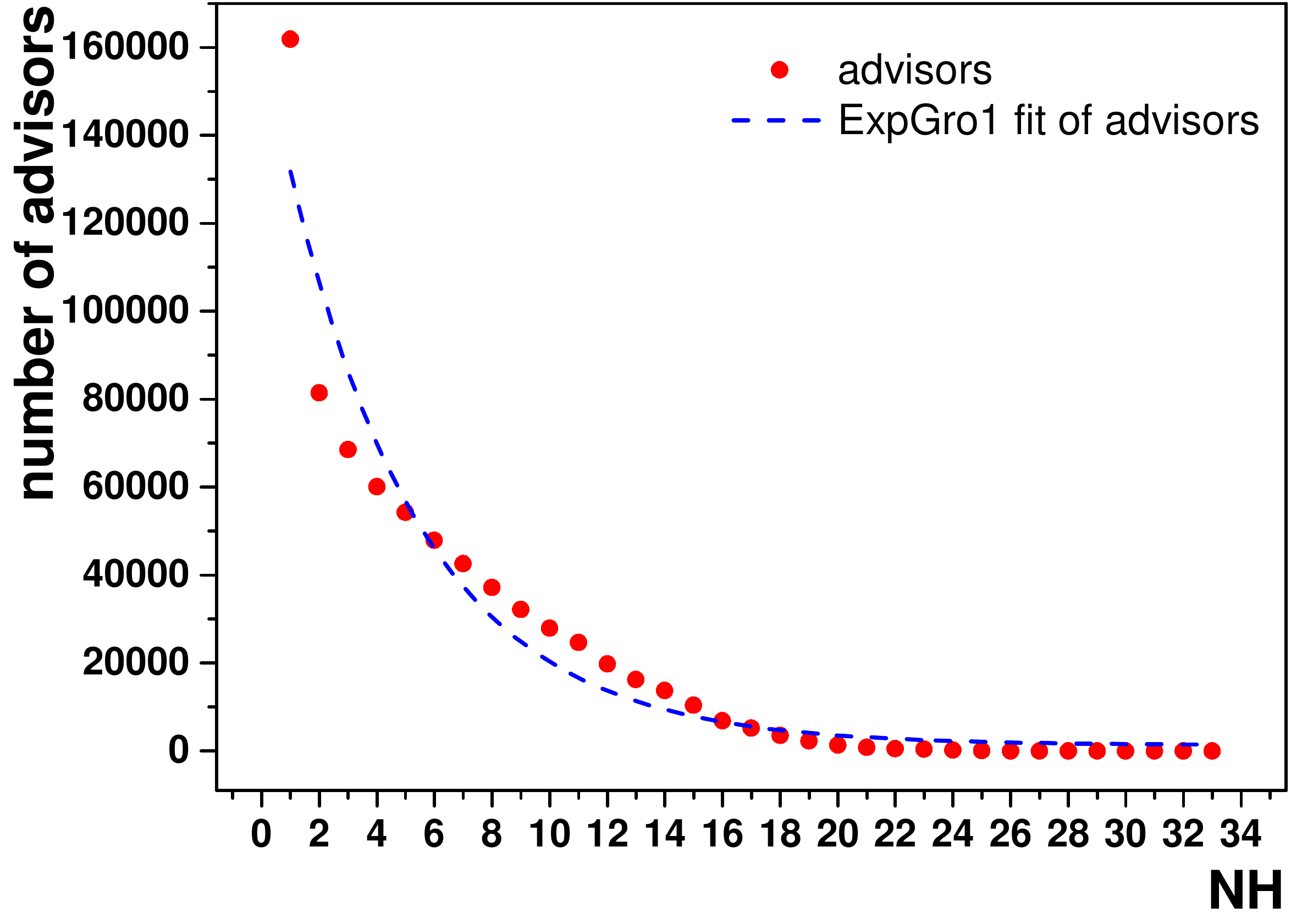}}
\caption{Fitted curves plot for $AA$, $NP$, $NC$, and $NH$. The horizontal axis represents (a) advisees' $AA$, (b) advisors' $AA$, (c) advisees' $NP$, (d) advisors' $NP$, (e) advisees' $NC$, (f) advisors' $NC$, (g) advisees' $NH$, and (h) advisors' $NH$, respectively. The vertical axis represents the number of scholars.}
\label{fig:2}
\end{figure}

Graphs in Fig. \ref{fig:2} are the results obtained from fitting methods for Fig. \ref{fig:1}. Specific fitting functions are summarised in Table \ref{tab:2}. In Table \ref{tab:2}, $y$ represents the number of advisees/advisors and $x$ represents $AA$/$NP$/$NC$/$NH$. In order to carry out effective fitting, we use different exponential functions to determine the relationship. In fact, the existing literature~\citep{newman2001structure} has pointed out that the degree distribution of the scientists' networks is between the exponential and the power-law. Obviously, our results are in consistent with aforementioned law.

\subsection{Correlation between advisors' academic characteristics and advisees' academic performance}
\begin{figure}[htbp]
\centering
\subfigure[Academic ages]{
\label{fig:5-a}
\includegraphics[width=0.48\textwidth]{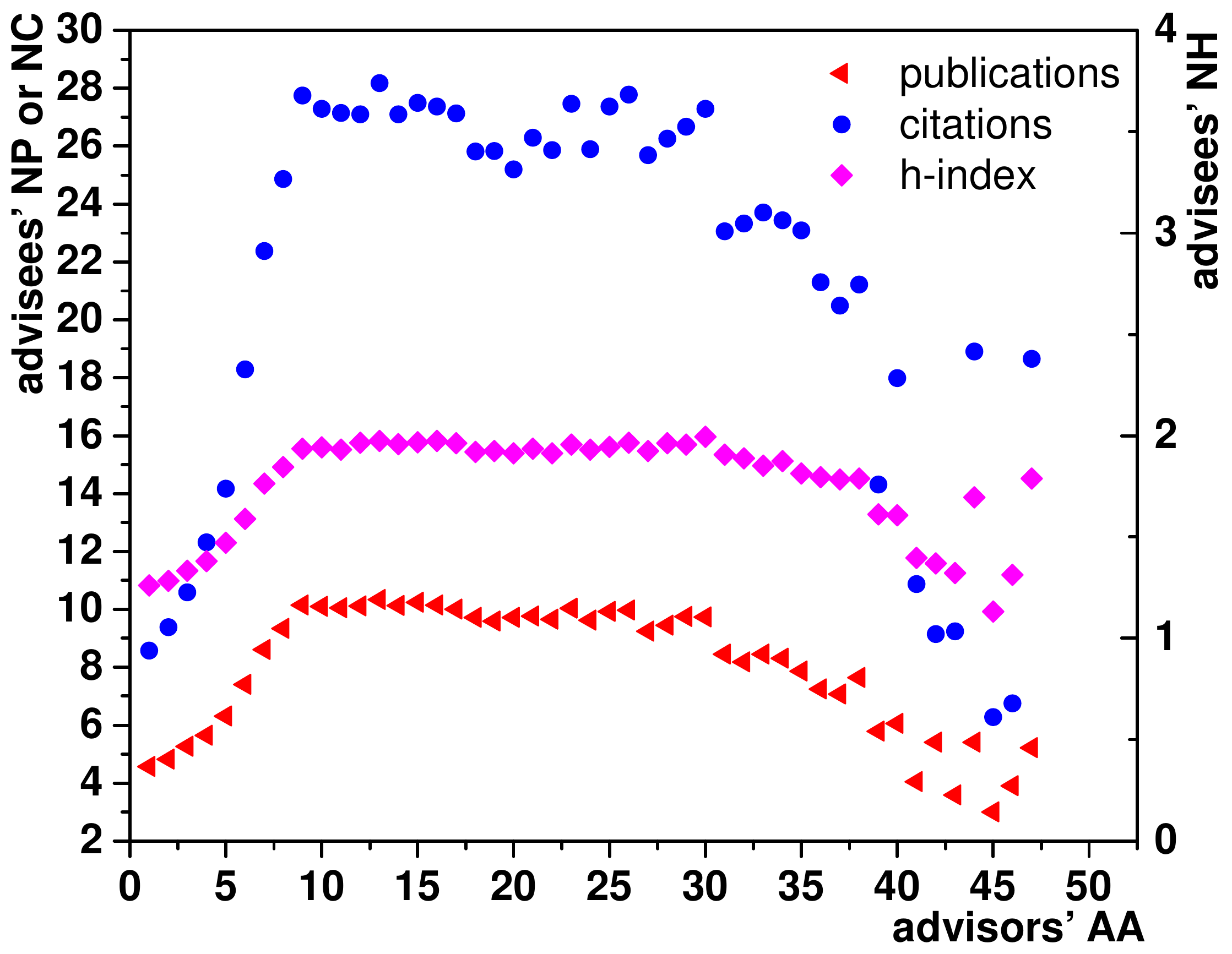}}
\subfigure[H-indices]{
\label{fig:5-d}
\includegraphics[width=0.48\textwidth]{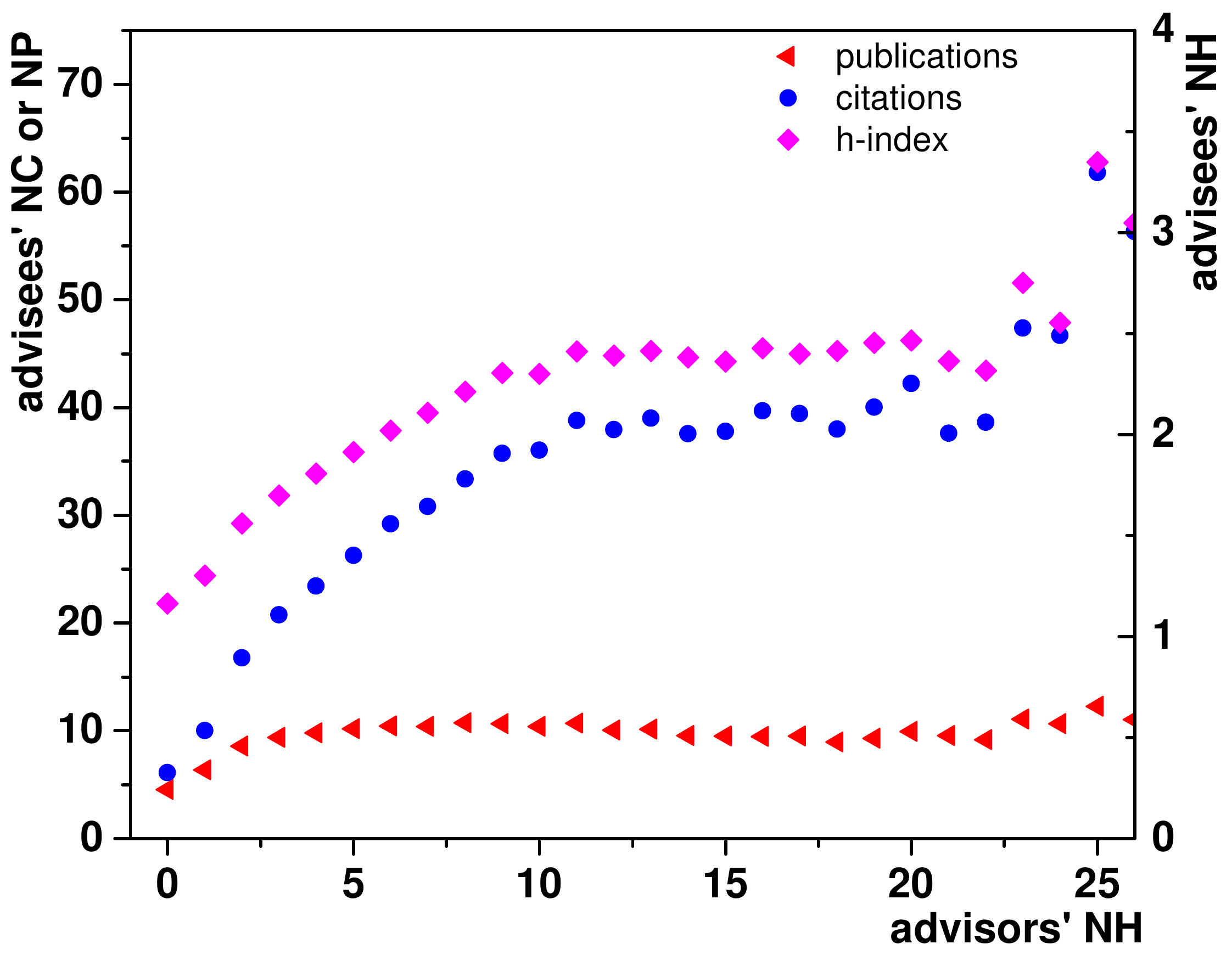}}
\caption{Relationship between the advisors' academic characteristics and advisees' academic performance for 10 years since collaborating with their advisors. The horizontal axis represents (a) advisors' academic age, (b) h-indices. In the subgraphs, the left vertical axis represents advisees' average number of publications and citations. The right vertical axis represents their average h-index.}
\label{fig:5}
\end{figure}

Fig. \ref{fig:5} shows the relationship between the advisors' academic characteristics and advisees' academic productivity as well as their impact. Academic productivity can be represented by the number of publications authored by one scholar, and academic impact can be approximated by a scholar's h-index~\citep{azoulay2012research,owens2013research}. It is a comprehensive indicator to evaluate scholars. Fig. \ref{fig:5-a} shows the correlation between advisors' $AA$ and average $NP$, $NC$, and $NH$ of their advisees 10 years after first collaborating with their advisors. It can be seen that the indicators have the same trend with the increasing of $AA$ because these indicators have a certain dependence. All of them experience an initial growth, then remain stationary, and finally reach a decline phase. When advisors' academic ages are in the range 1-9, advisees' academic performance shows a linear growth trend. When advisors' academic age reaches 28, advisees' performance shows a downward trend. In the period when advisors' academic age is between 10 and 27, advisees' average achievements reach the highest level and basically remain unchanged.

It is interesting to consider the possible factors underlying this phenomenon. Prior studies have noted the importance of advisors' career coaching and social support in advisees' development. The current study~\citep{Wang2017Scientific} finds that the productivity of scholars is dynamic over the course of their own scientific career. For scholars whose $AA < 12$, the annual productivity rises up slowly. If their $AA > 24$, their productivity will decline slightly with the $AA$. In the range of $12 < AA < 24$, their annual productivity are nearly fixed. Advisees' average productivity/impact shows a linear growth trend because their advisors' productivity in the range $1 \leq AA \leq 9$ increases linearly. This reflects that the relationship between advisors' career phases and advisees' outcomes in terms of productivity and impact actually exists in the scientific mentorship. Our analyses indicate the relationship between collaborations with different $AA$ advisors and the future career performance of advisees thus need deeper exploration.

Here are some possible reasons for our findings: (a) Academic research is a long-term accumulation process. Junior advisors need time to accumulate experience in conducting research and mentoring advisees. Compared with senior ones, their capacity and resources are limited. Senior advisors have cumulative advantages, both in academic performance and resources. (b) When the academic age of advisors is at the intermediate level (in the range of 10-27 years), their academic capability may reach a certain level as well. With sufficient resources, they can provide their advisees with the most helpful guidance. In addition to that, they are also willing to strive to instruct their advisees, because their own academic performance can be improved in this way. (c) Few senior advisors have a relatively long academic career (as shown in Fig. \ref{fig:1-a}), that's why senior advisors of academic age above 35 have poor performing advisees. The random fluctuation may cause this phenomenon, and thus further study is required to untangle the underlying reasons. In general, the correlation between advisees' high performance and advisors' academic ages is certainly reasonable, but it is not the causation of scholars' success. Hard work and passion are still essential to scientists' success in the scientific career.

Similarly, in Fig. \ref{fig:5-d}, the abscissa represents the advisors' $NH$. In this subgraph, the left vertical axis represents advisees' average $NP$, $NC$ and right axis represents their average $NH$. Comparing with advisees' $NC$ and $NH$, advisees' $NP$ has the least obvious trend with the increase of advisors' $NH$. 
The phenomenon elicits the truth that quantity is not equal to quality especially in academia. The maximum value of advisors' h-index we have considered is 25 because there are few advisors with high $NH$ values (see Fig. \ref{fig:1}). The value of indicators for advisees' fluctuates widely along with the growth of abscissa. So the data is binned in order to make the trend more clear.

\subsection{An accomplished advisor brings up a skilled advisee}

\begin{figure}[htbp]
\centering
\subfigure[5]{
\label{fig:6-a}
\includegraphics[width=0.47\textwidth]{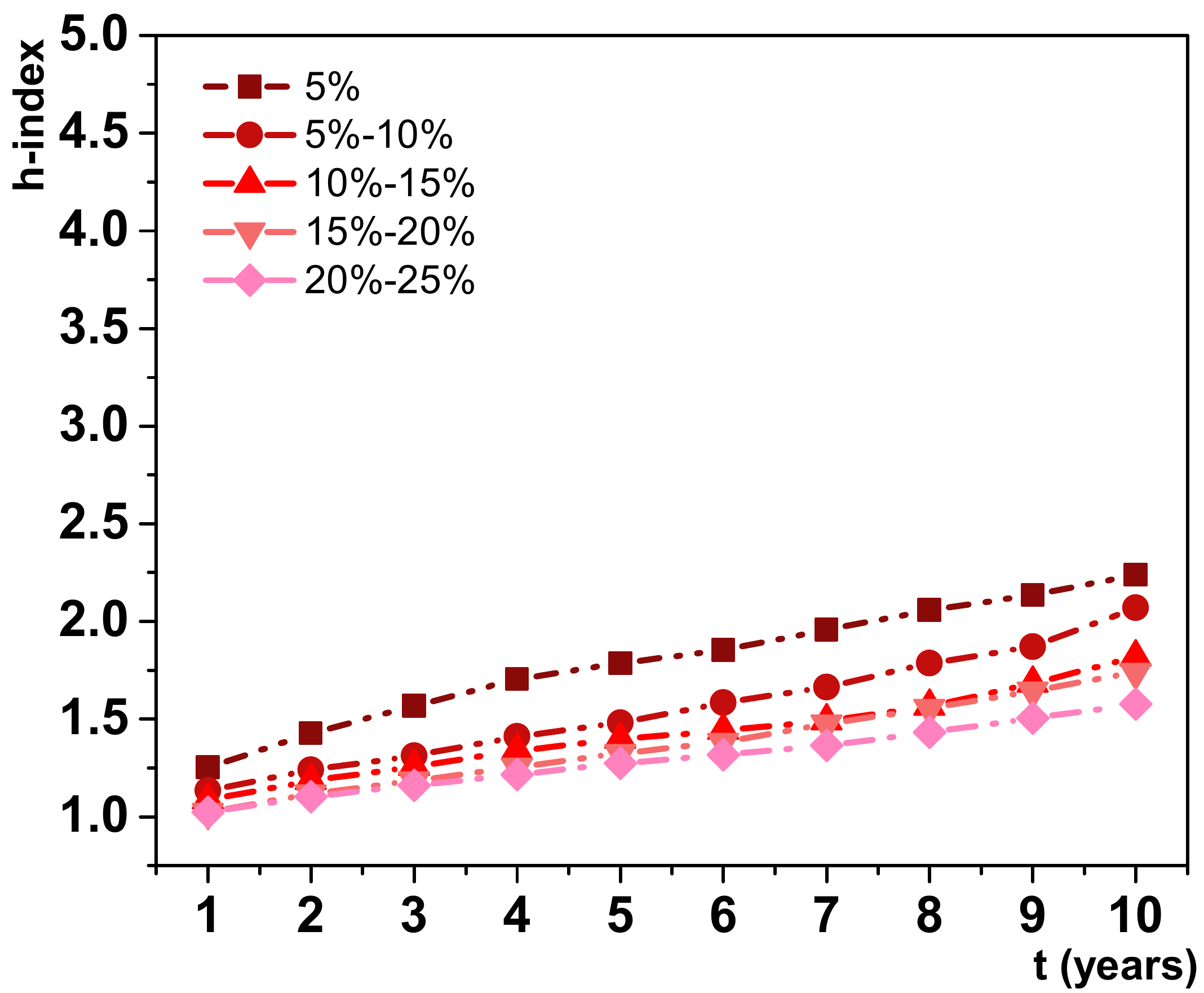}}
\subfigure[10]{
\label{fig:6-b}
\includegraphics[width=0.47\textwidth]{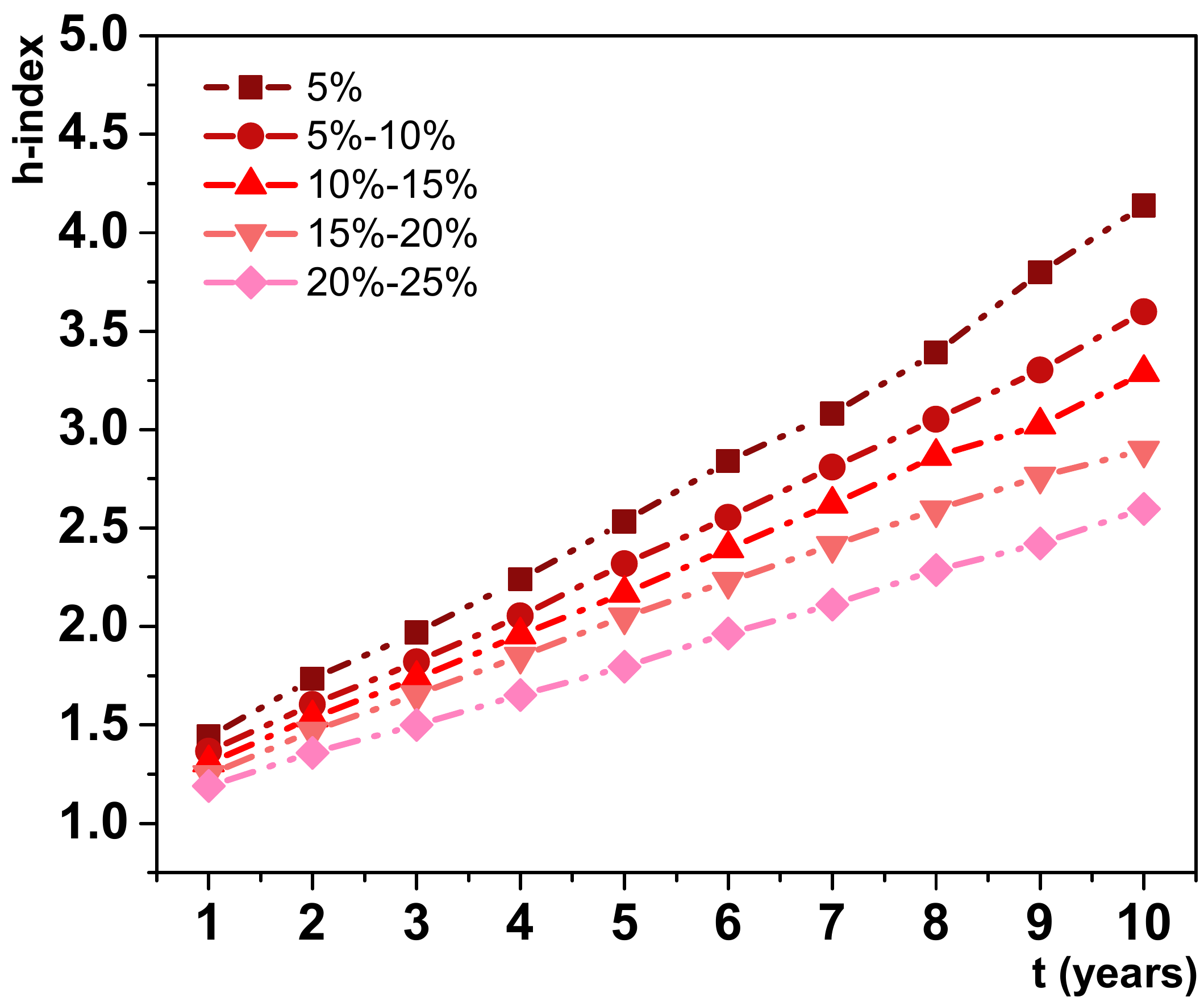}}\\
\subfigure[15]{
\label{fig:6-c}
\includegraphics[width=0.47\textwidth]{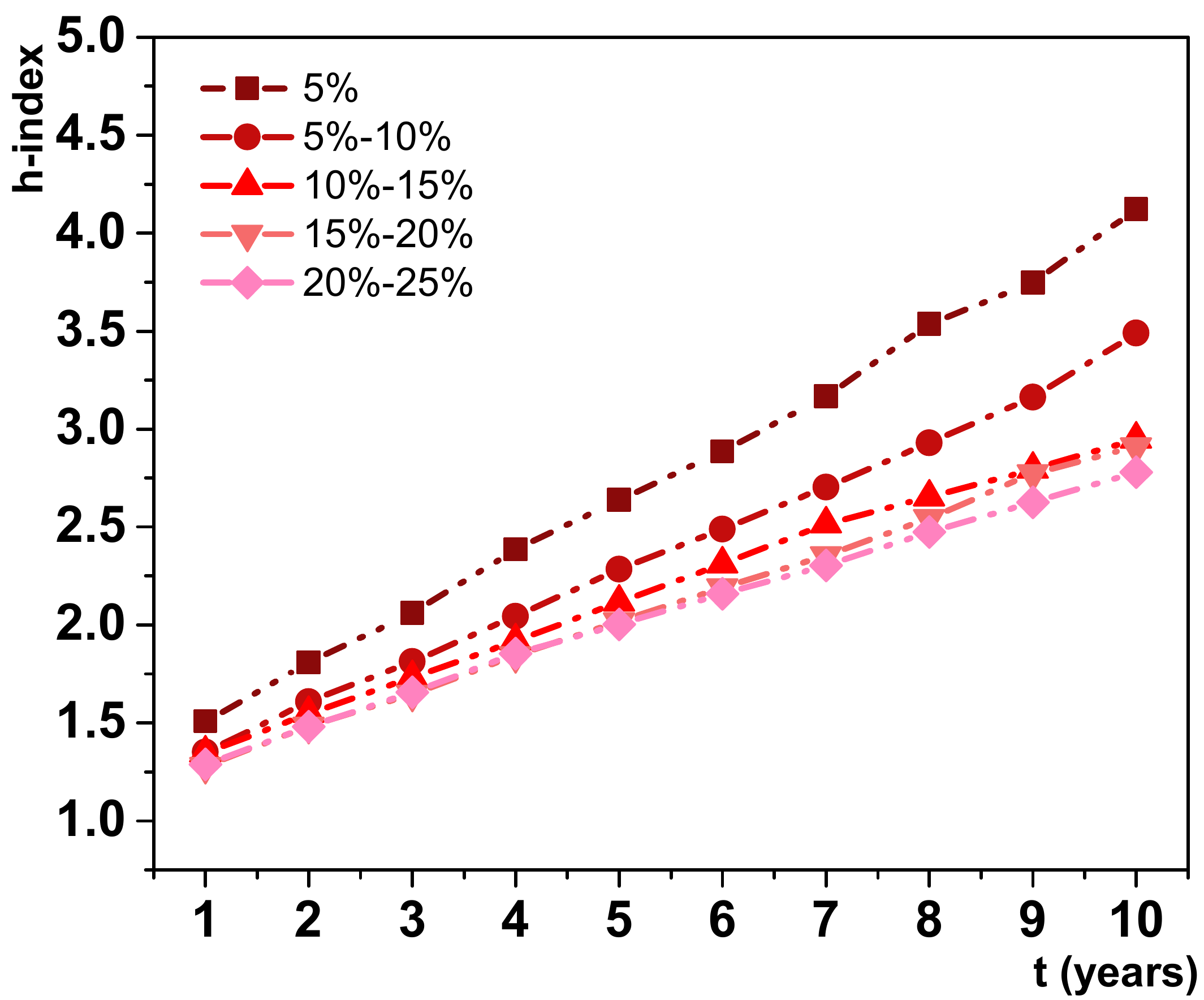}}
\subfigure[20]{
\label{fig:6-d}
\includegraphics[width=0.47\textwidth]{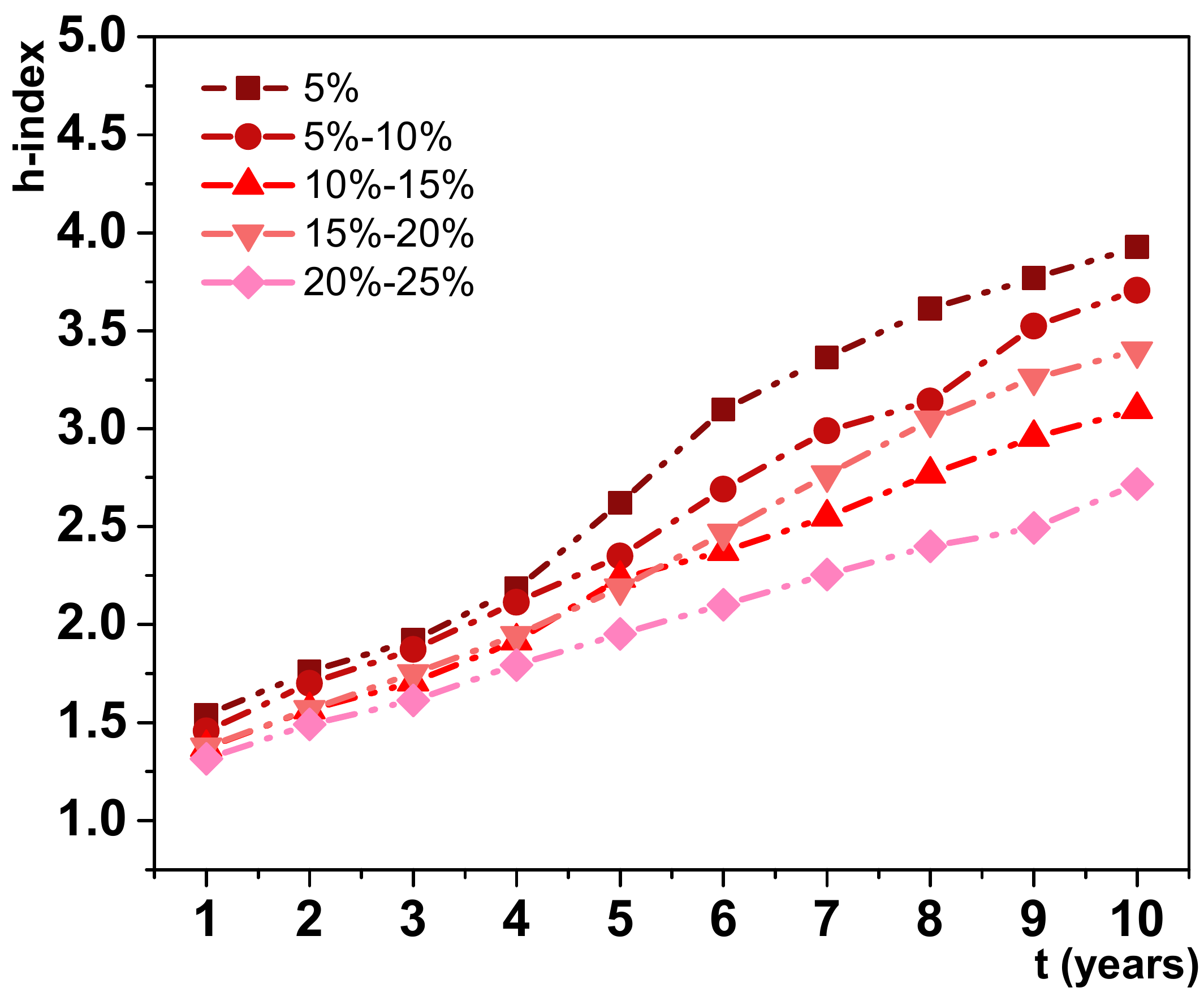}}\\
\subfigure[25]{
\label{fig:6-e}
\includegraphics[width=0.47\textwidth]{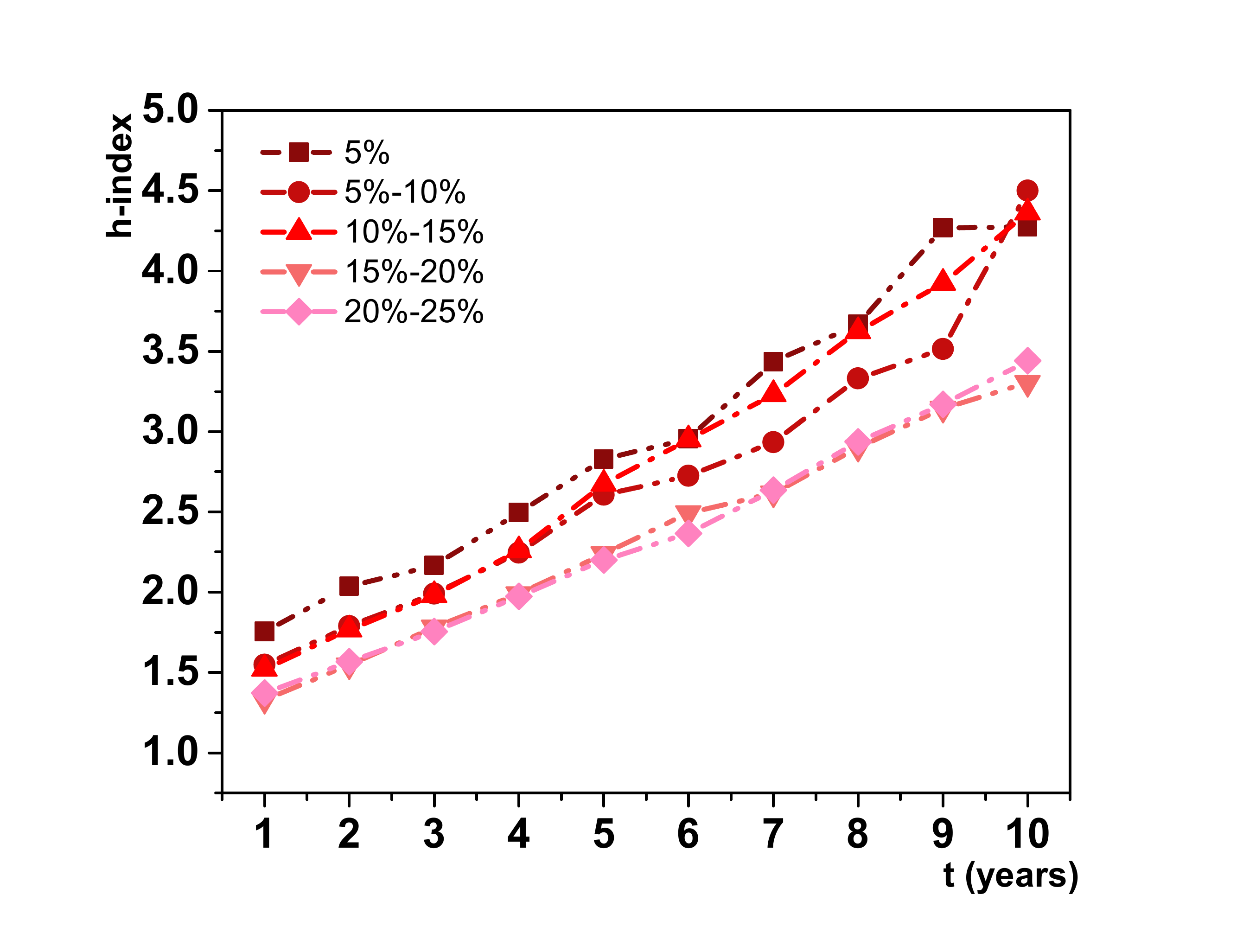}}
\caption{Advisees' h-indices whose advisors' academic ages are (a) 5, (b) 10, (c) 15, (d) 20, (e) 25, as well as their h-indices are ranking at the top 5\%, 5\%-10\%, 10\%-15\%, 15\%-20\%, 20\%-25\%. The polylines represent the trends of all advisees' average h-index over time.}
\label{fig:6}
\end{figure}

According to the influence of advisors' academic ages on advisees, Fig. \ref{fig:6} indicates advisees' h-indices whose advisors' academic ages are 5, 10, 15, 20, 25, and the ranking of their h-indices begins at the top 5\%, 5\%-10\%, 10\%-15\%, 15\%-20\%, 20\%-25\%, respectively.
The polylines represent the trends of all advisees' average h-index over time. It can be seen from the data in Fig. \ref{fig:6}, advisors with different rankings in h-index can bring their advisees different h-indices. It is commonly perceived that high-achieving advisors literally cause their advisees higher academic productivity and impact. Selective bias is one possible reason as top advisors tend to select quality students. It is interesting to observe that as the advisors' academic age rises, the gap among advisees narrows.

\begin{figure}[htbp]
\centering
\subfigure[Number of publications]{
\label{fig:7-a}
\includegraphics[width=0.47\textwidth]{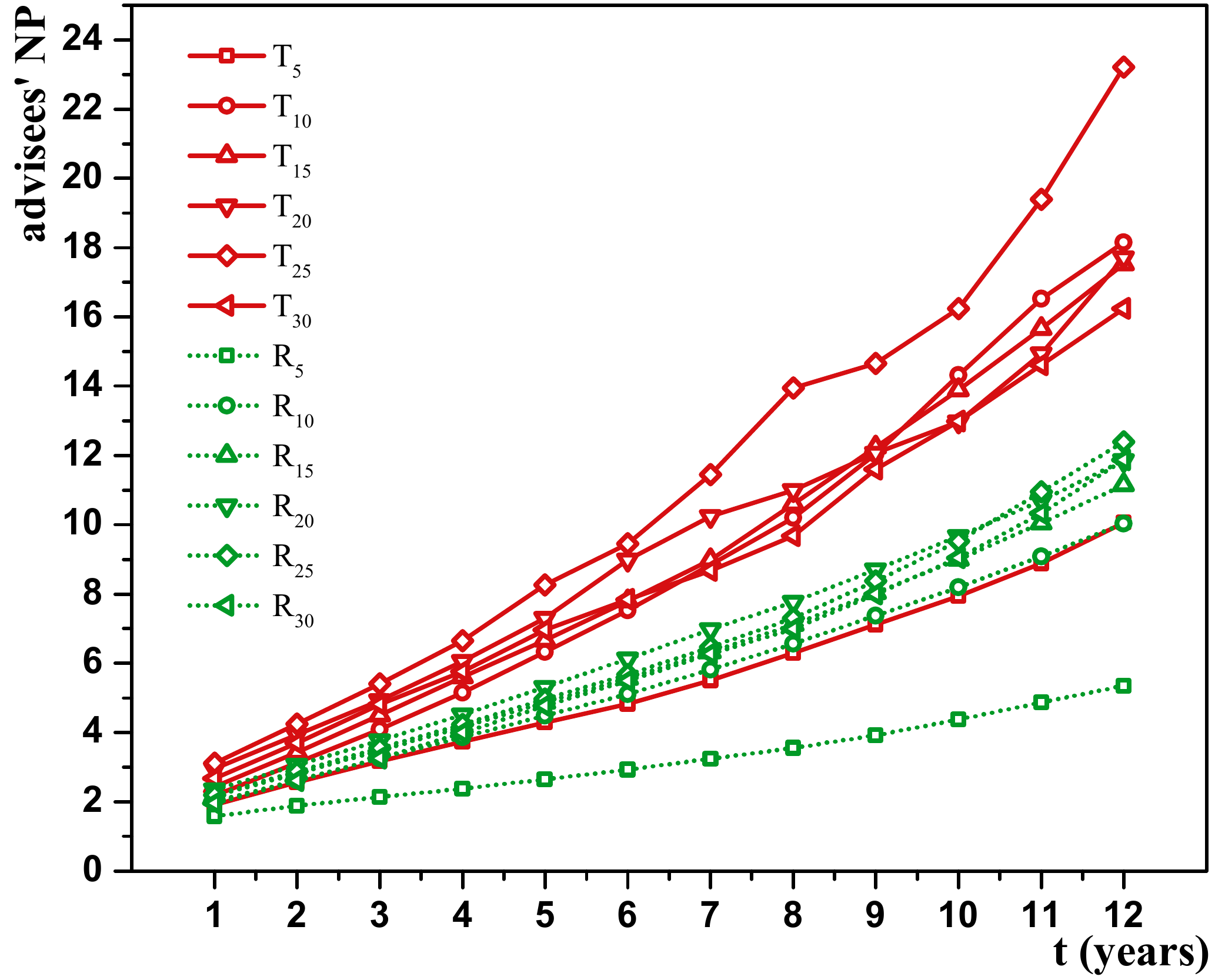}}
\subfigure[Publications' difference]{
\label{fig:7-d}
\includegraphics[width=0.47\textwidth]{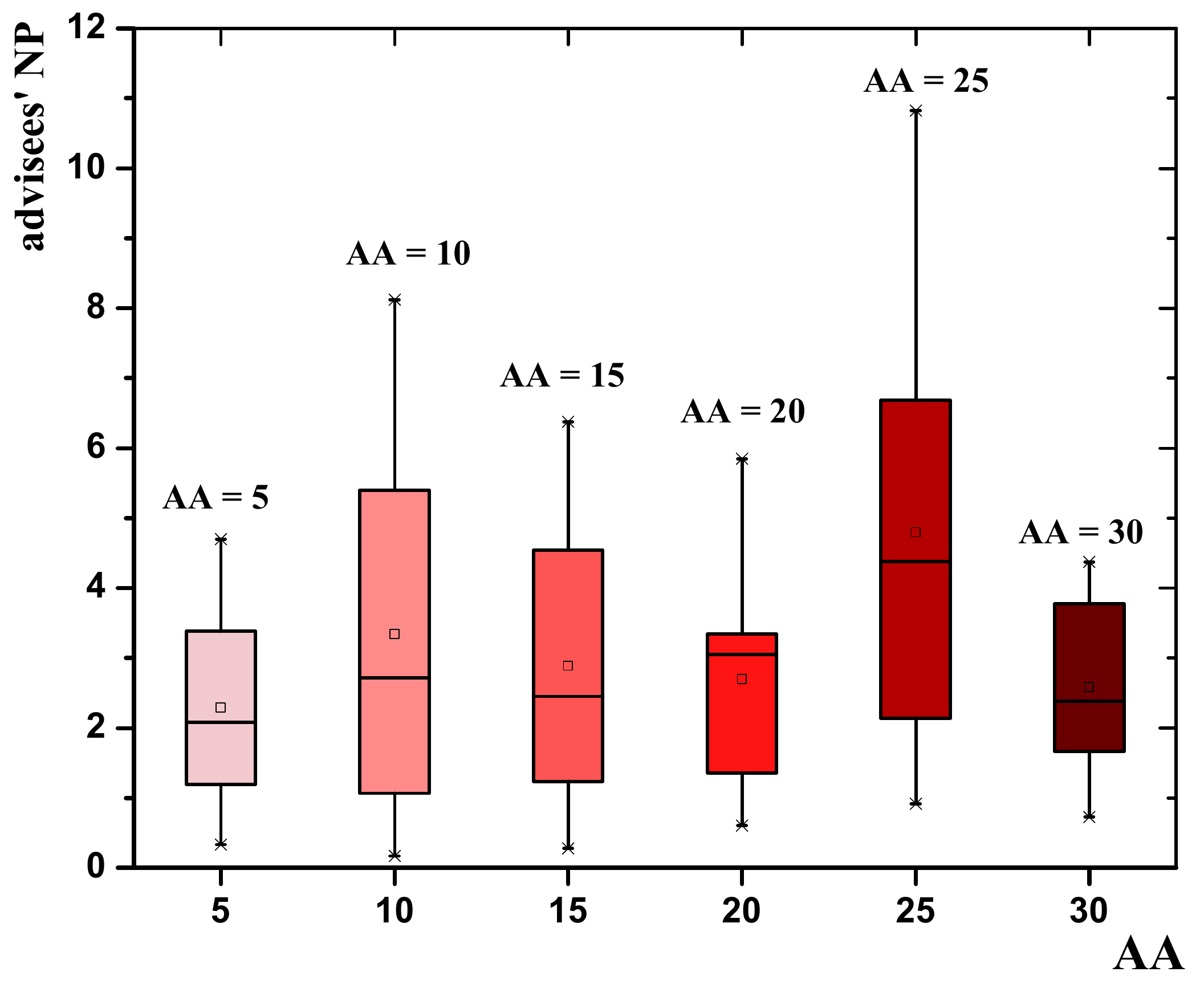}}\\
\subfigure[Citations]{
\label{fig:7-b}
\includegraphics[width=0.47\textwidth]{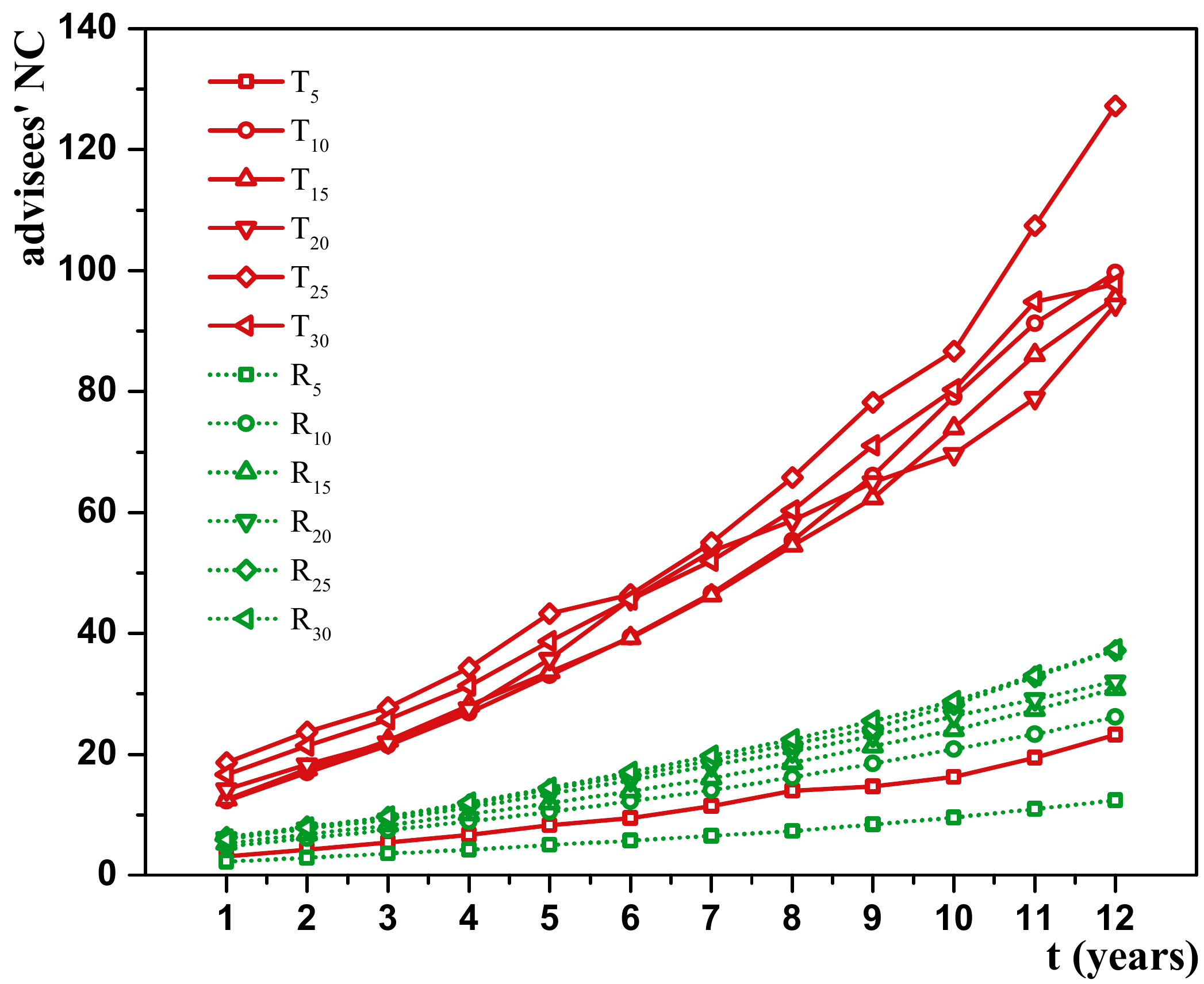}}
\subfigure[Citations' difference]{
\label{fig:7-e}
\includegraphics[width=0.47\textwidth]{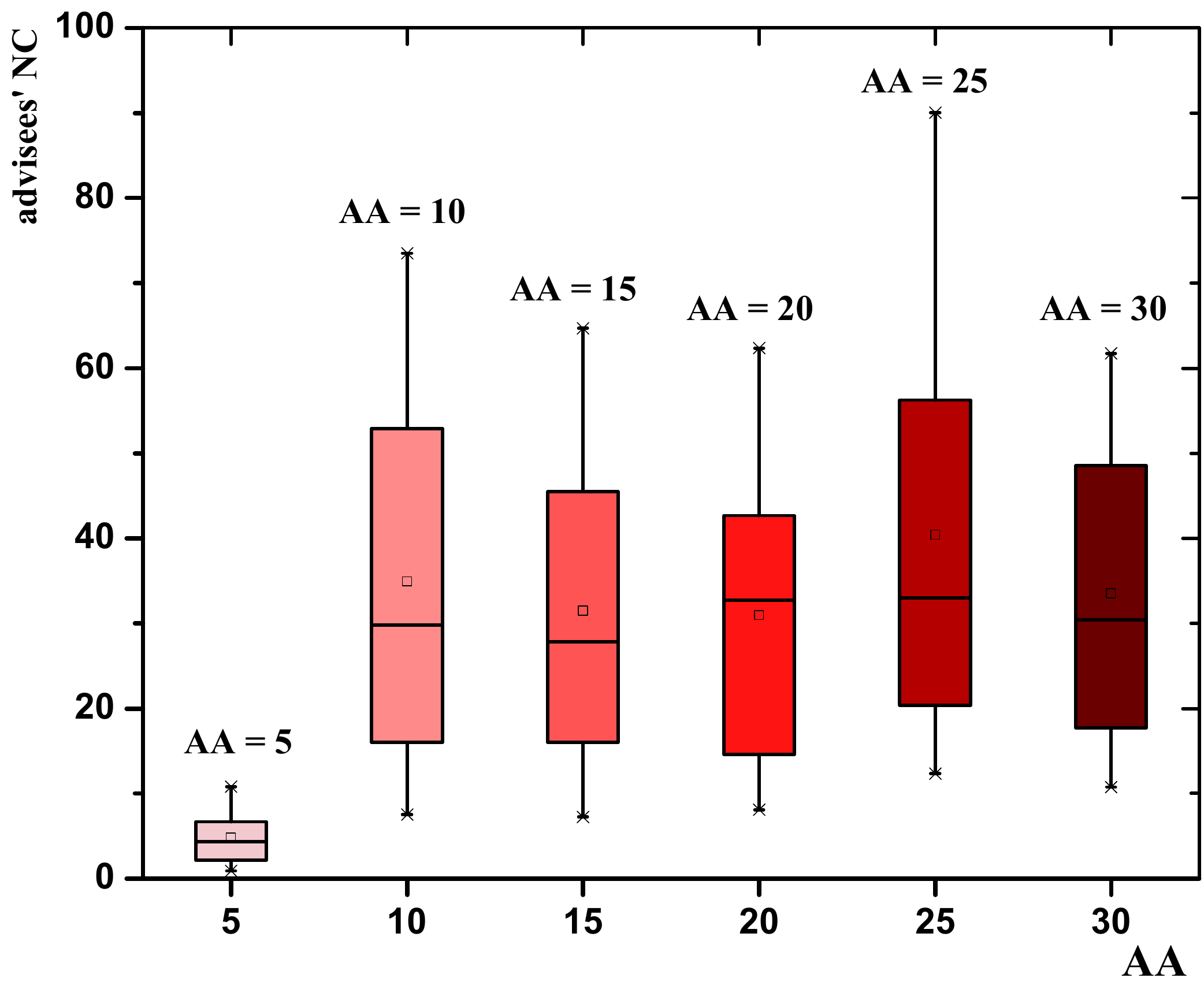}}\\
\subfigure[H-indices]{
\label{fig:7-c}
\includegraphics[width=0.46\textwidth]{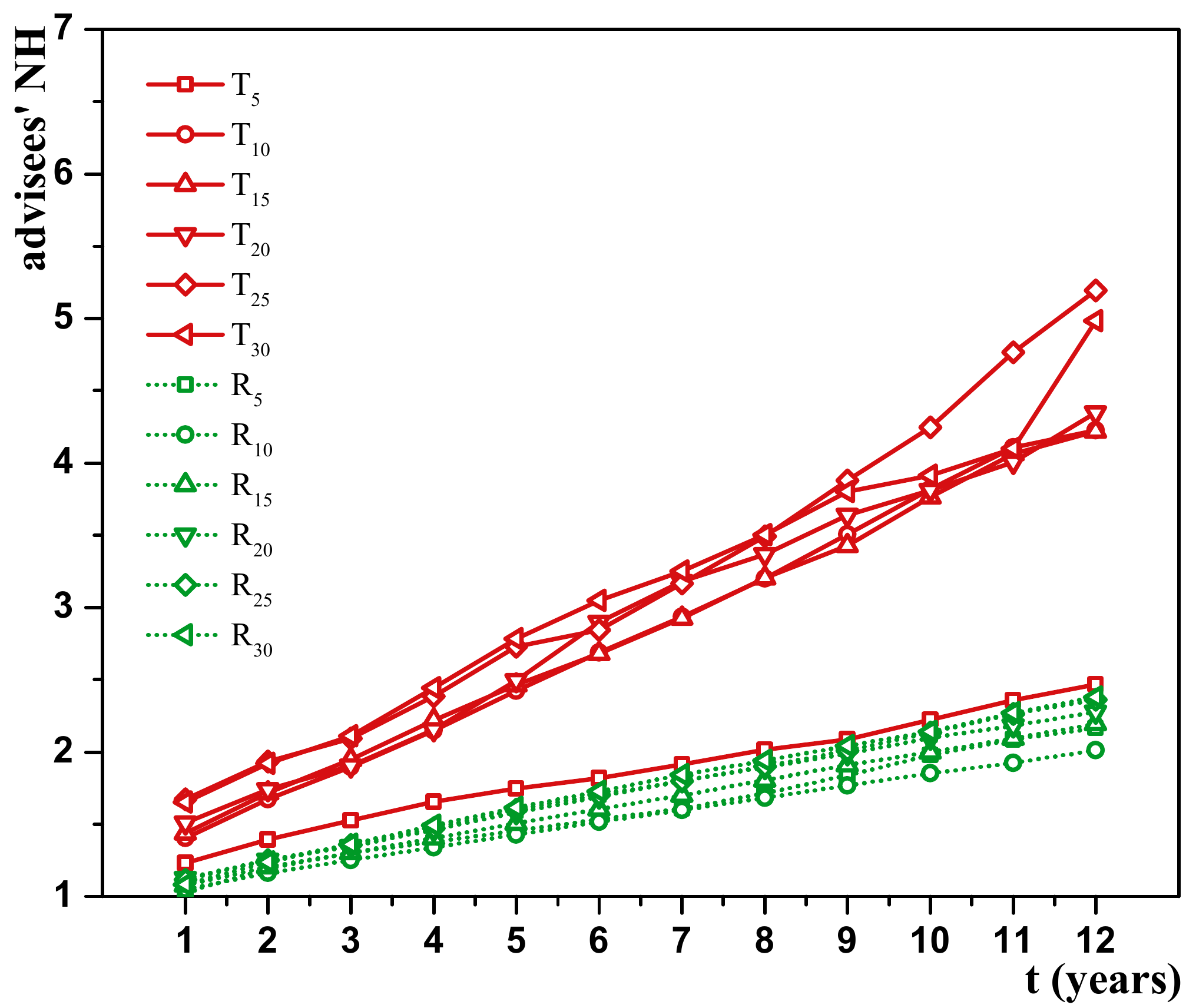}}
\subfigure[H-indices' difference]{
\label{fig:7-f}
\includegraphics[width=0.46\textwidth]{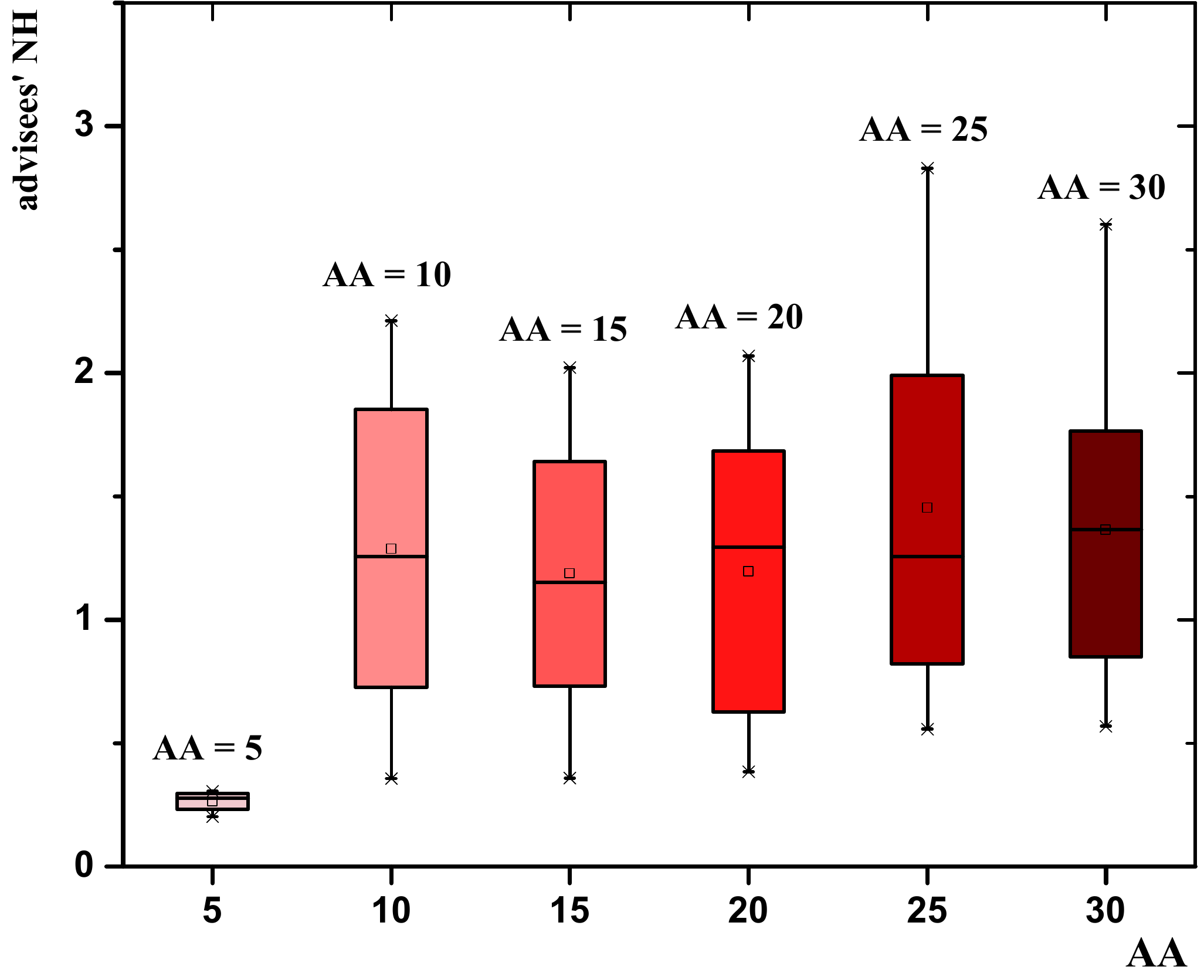}}
\caption{Average academic performance of all advisees supervised by $Top_{10}$ advisors and $Res$ advisors. $T_n$ corresponds to $Top_{10}$ advisors whose academic ages are $n$. $R_n$ corresponds to the rest whose academic ages are $n$. (a), (c), and (e) show the advisees' number of publications, citations and h-indices coached by different advisors, respectively. Details of their statistical mean and deviation are shown in (b), (d), and (f), respectively.}
\label{fig:7}
\end{figure}

The advisees supervised by advisors whose h-index rank the top 10\% always have the highest h-index. But it is not clear how significant the differences among these advisees are. On the basis of the definition of scholars' impact, here we regard the top ten percent of advisors in h-index as $Top_{10}$ advisors. $Res$ advisors represent the rest of advisors. We try to find out the differences between advisees mentored by $Top_{10}$ advisors or $Res$ advisors. We convert the problem from a qualitative analysis to the quantitative analysis. Fig. \ref{fig:7} reflects advisees' average $NP$, $NC$, and $NH$ over time since advisees first collaborated with their advisors. In Fig. \ref{fig:7}, advisors are differentiated with academic ages and h-indices. In the figures, $T_n$ corresponds to $Top_{10}$ advisors whose academic ages are $n$ while $R_n$ corresponds to $Res$ advisors whose academic ages are $n$. From the figures, we can see that advisors' $NP$ results in the least obvious discrepancy between advisees supervised by $Top_{10}$ advisors and $Res$ advisors. The most striking observation to emerge from the data comparison is the growth rate of each indicator. Advisees mentored by $Top_{10}$ advisors have higher $NP$, $NC$, and $NH$ than the others. Moreover, their growth rate is higher. This result may be explained by the Matthew Effect~\citep{langfeldt2015excellence}. In academia, while excellent scholars' accomplishments and reputation tend to snowball, those with modest accomplishments have greater difficulty to improve their impact.

\begin{figure}[htbp]
\centering
\subfigure[Number of publications]{
\label{fig:8-a}
\includegraphics[width=0.47\textwidth]{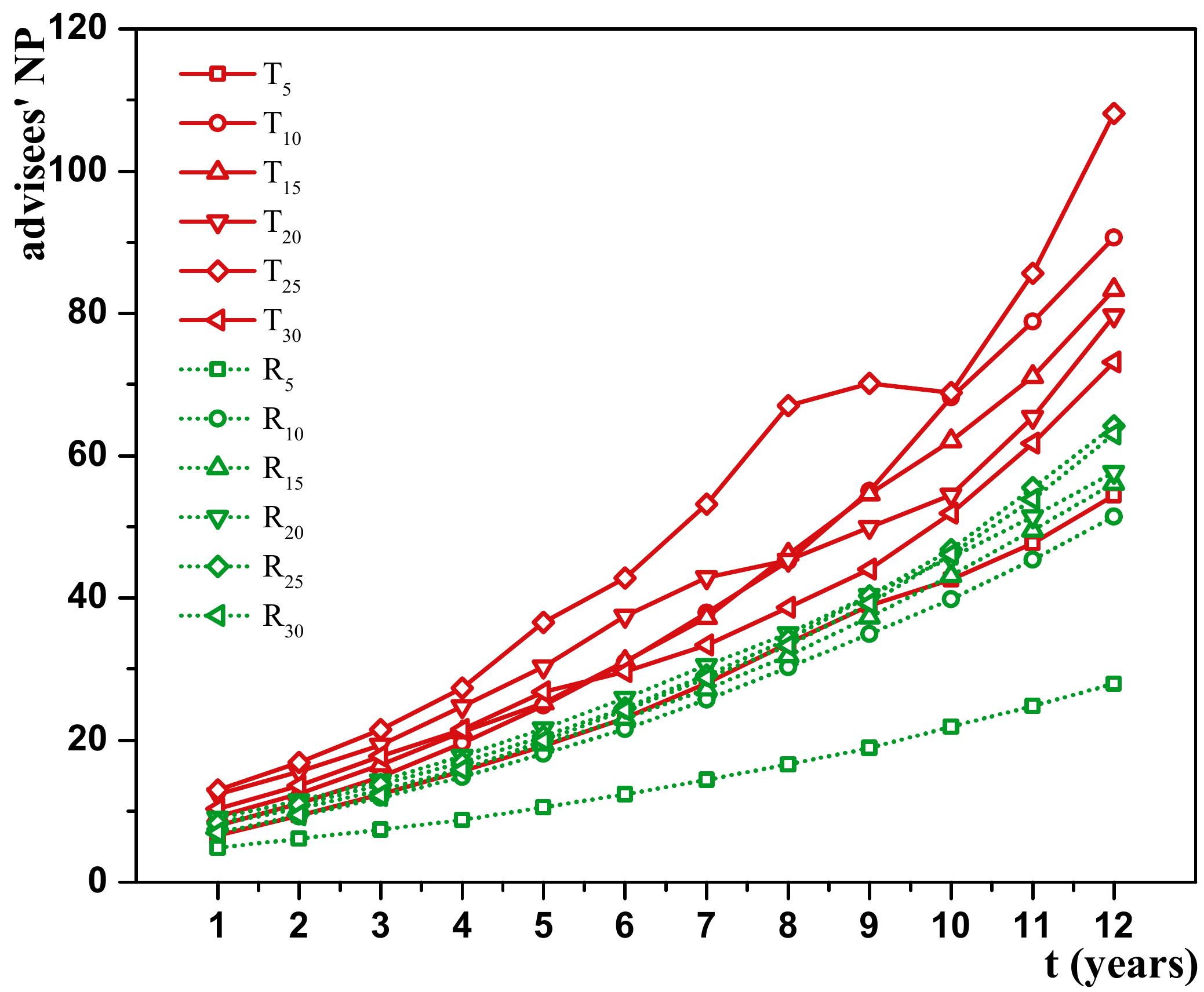}}
\subfigure[Publications' difference]{
\label{fig:8-b}
\includegraphics[width=0.47\textwidth]{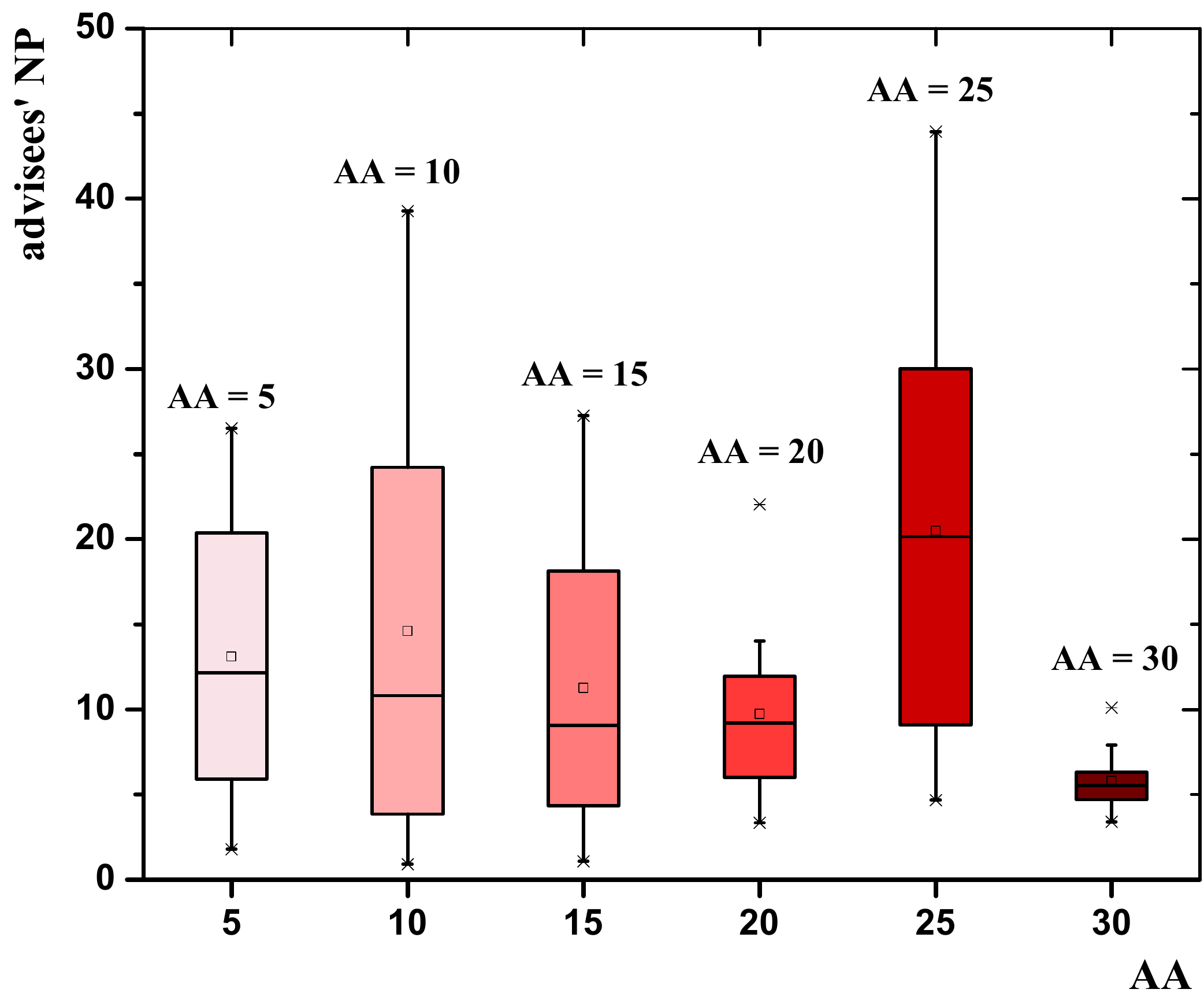}}\\
\subfigure[Citations]{
\label{fig:8-c}
\includegraphics[width=0.47\textwidth]{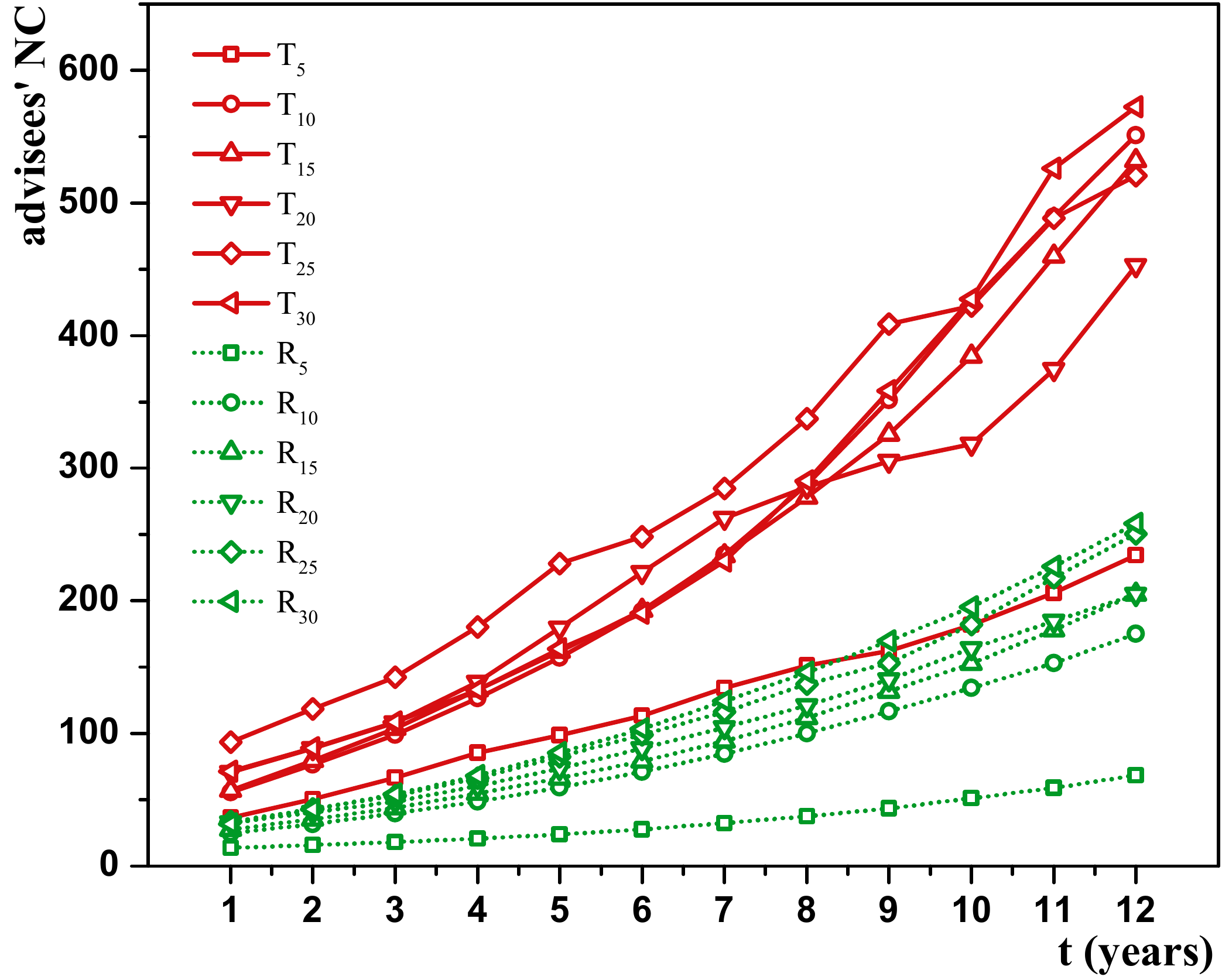}}
\subfigure[Citations' difference]{
\label{fig:8-d}
\includegraphics[width=0.47\textwidth]{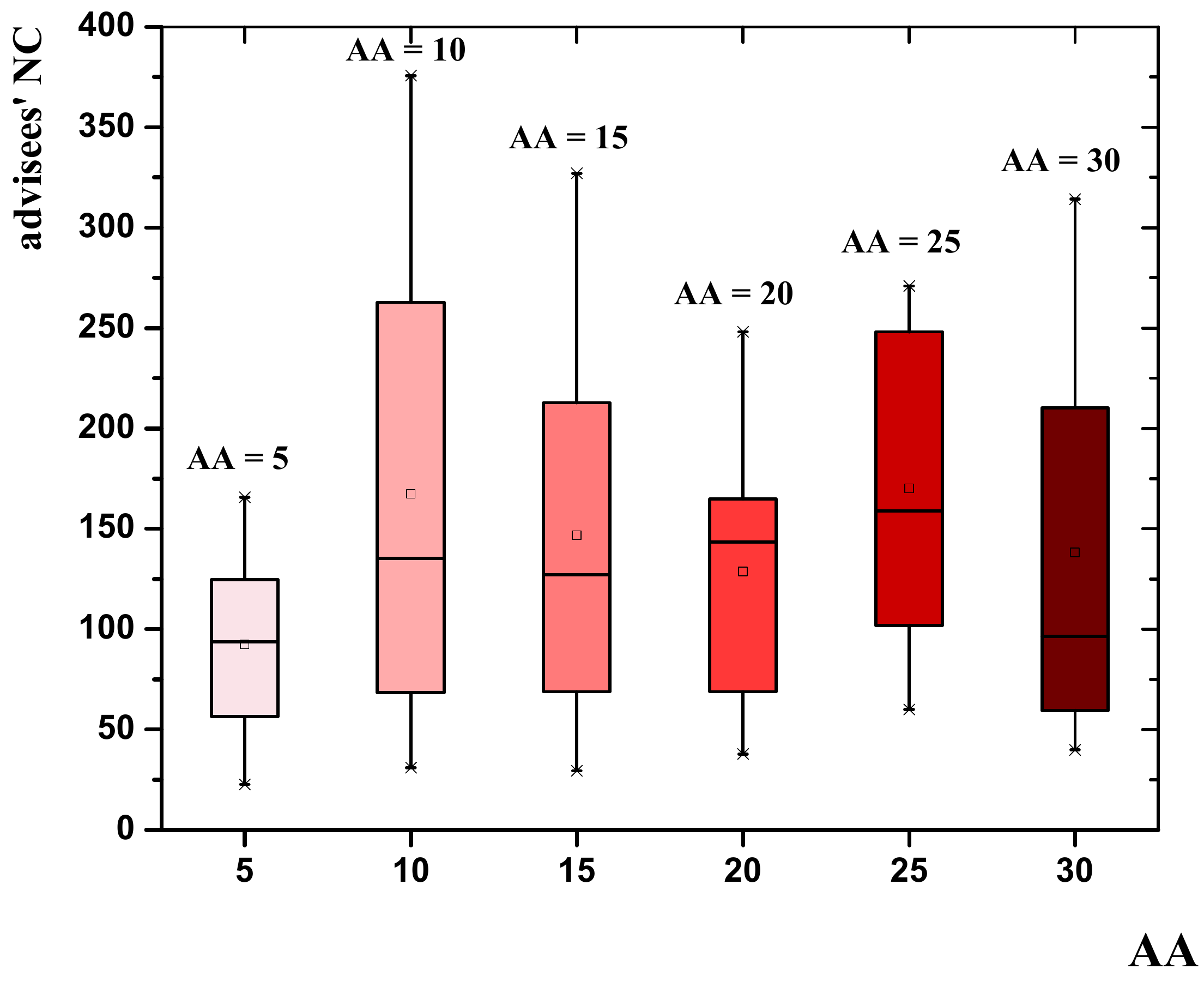}}\\
\subfigure[H-indices]{
\label{fig:8-e}
\includegraphics[width=0.465\textwidth]{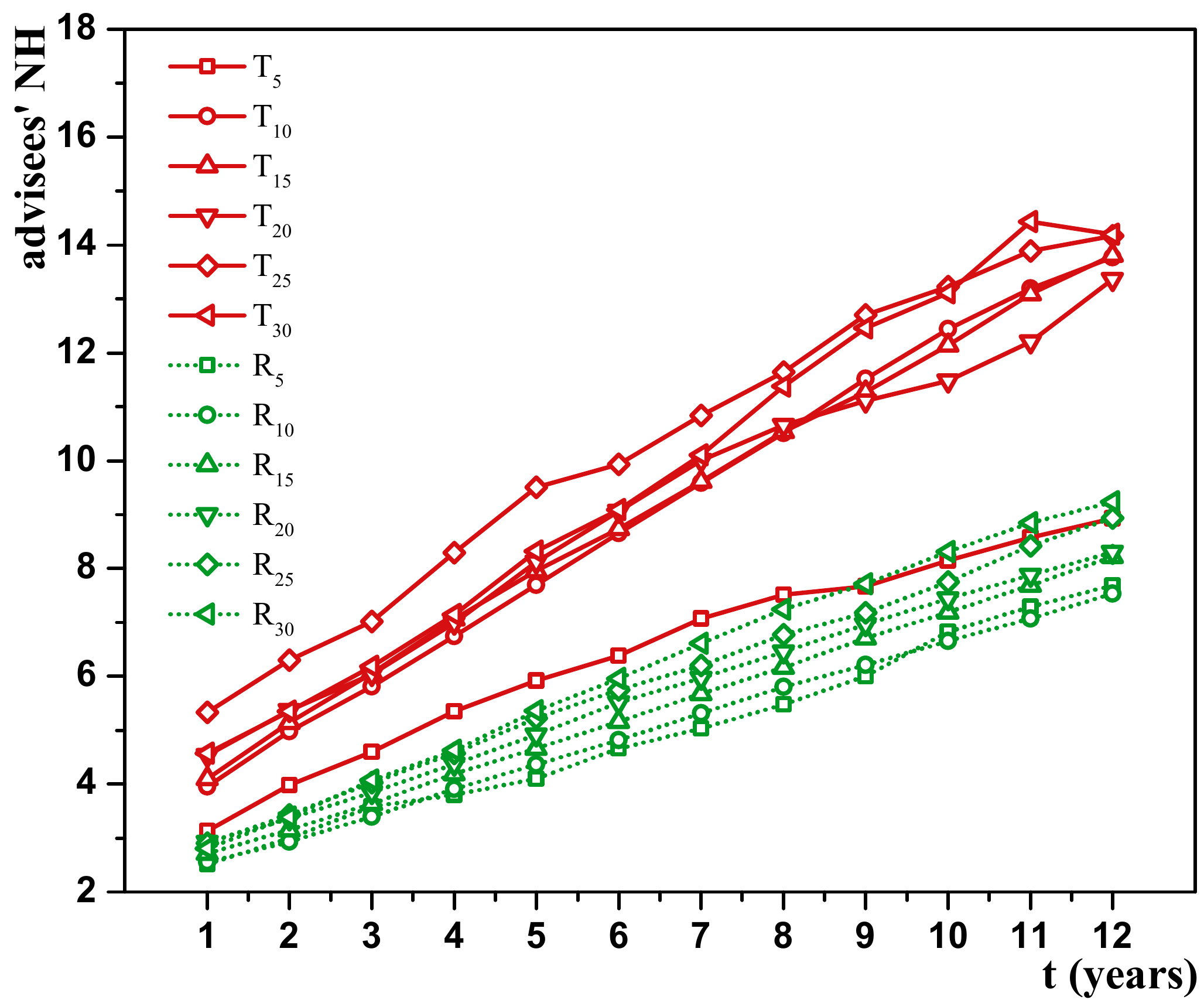}}
\subfigure[H-indices' difference]{
\label{fig:8-f}
\includegraphics[width=0.46\textwidth]{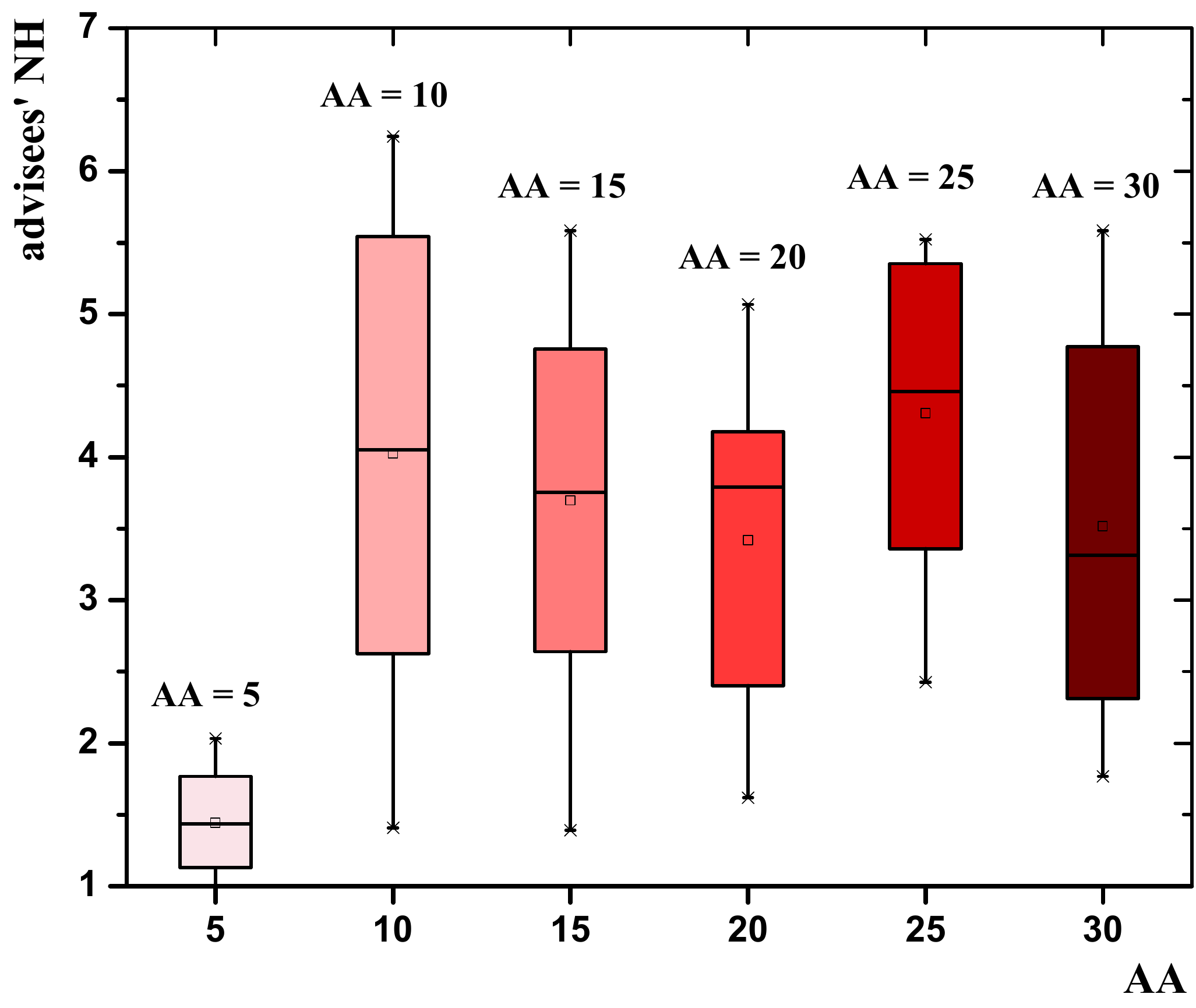}}
\caption{$Top_{10}$ advisees' academic performance supervised by $Top_{10}$ advisors and $Res$ advisors.  (a), (c), and (e) show the $Top_{10}$ advisees' number of publications, citations and h-indices coached by different advisors. Differences between
$Top_{10}$ advisees' $NP$, $NC$, and $NH$ who are mentored by $Top_{10}$ advisors or $Res$ advisors are highlighted in (b), (d), and (f).}
\label{fig:8}
\end{figure}

In Fig. \ref{fig:8}, similarly, which is consistent with assessment standards of advisors, the advisees whose h-index rank the top 10\% of each indicator are considered as the $Top_{10}$ advisees. The differences between $Top_{10}$ advisees' $NP$, $NC$, and $NH$, who are mentored by $Top_{10}$ advisors or $Res$ advisors are highlighted in Fig. \ref{fig:8-d}, Fig. \ref{fig:8-e}, and Fig. \ref{fig:8-f}. To the same extent as the previous results, comparing with $NC$ and $NH$, the discrepancy between advisees' average $NP$ mentored by $Top_{10}$  advisors or $Res$ advisors is not obvious. However, considering academic ages of the advisors, $Top_{10}$ advisees mentored by $Top_{10}$ advisors whose academic age is 30 have the lowest average $NH$. For $Res$ advisors, the same scenario occurs to advisors whose academic ages are 5. It is different from advisors' $NC$ and $NH$. This phenomenon gives us inspiration that only considering the number of publications to evaluate scholars is insufficient because quantity is not always equal to quality. As a result, a number of comprehensive indices have appeared to evaluate scholars. In Fig. \ref{fig:7} and Fig. \ref{fig:8}, it can be seen from the subgraphs that comparing with all of the advisees, $Top_{10}$ advisees mentored by $Top_{10}$ advisors or $Res$ advisors make more distinct values in the average $NP$, $NC$, and $NH$.

\begin{figure}[htbp]
\centering
\subfigure[Proportion]{
\label{fig:9-a}
\includegraphics[width=\textwidth]{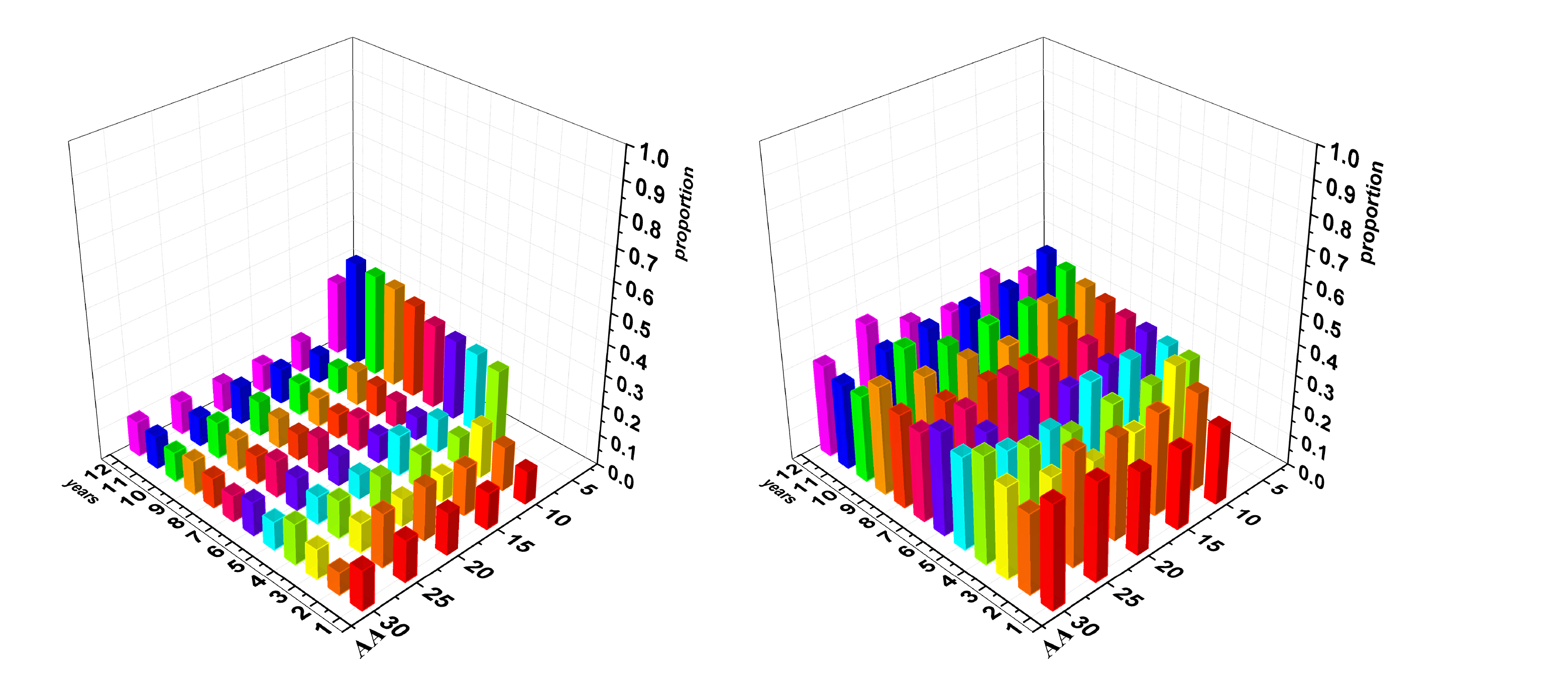}}
\subfigure[Proportion differences]{
\label{fig:9-b}
\includegraphics[width=0.47\textwidth]{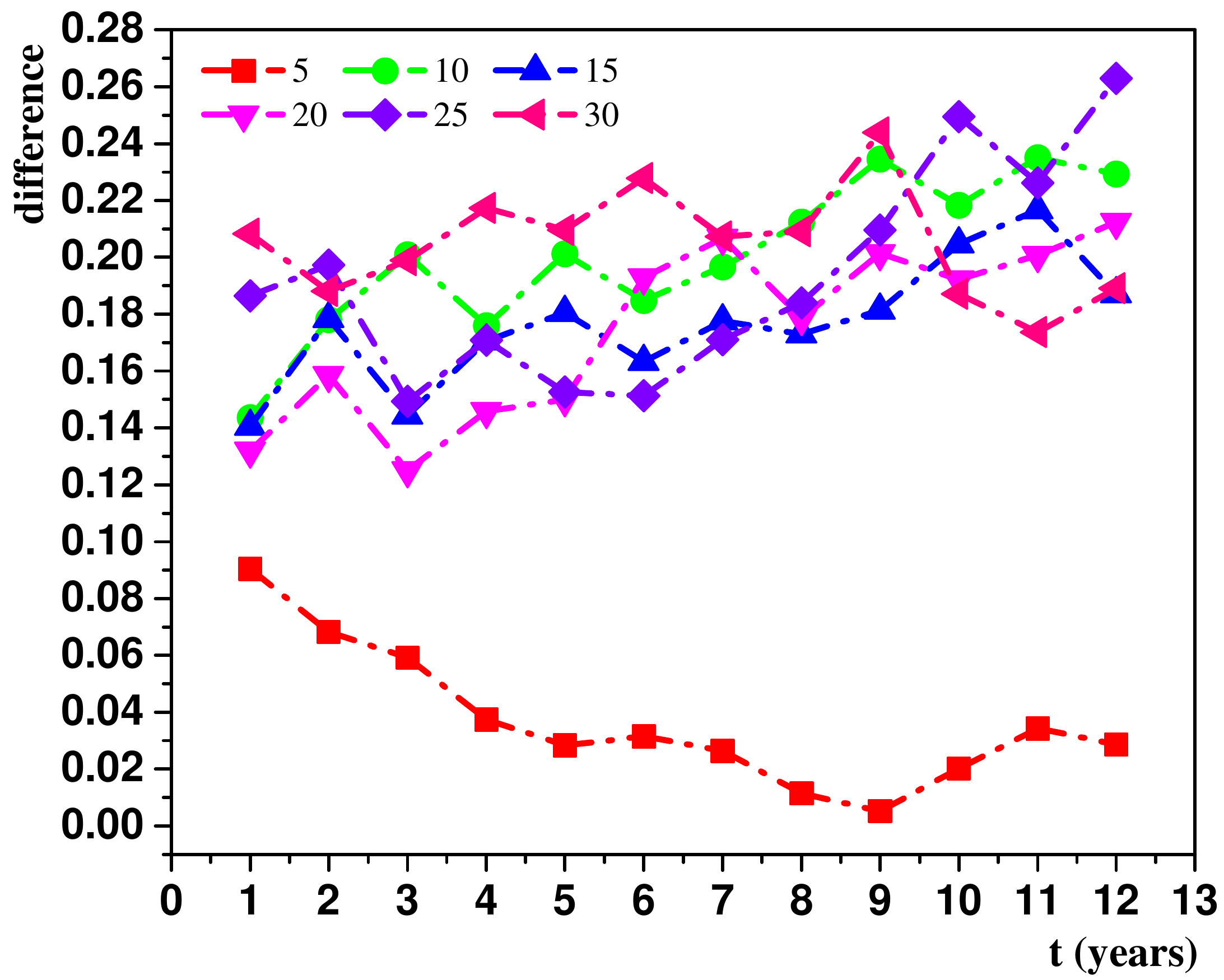}}
\caption{Correlations among advisor-advisee's collaboration duration, advisors' academic age, the proportion of advisees' h-index ranking the top 10\%. (a) Advisees' probability of h-index ranking the top 10\% mentored by $Res$ and $Top_{10}$ advisors. (b) The proportion of differences between advisees mentored by $Top_{10}$ advisors and $Res$ advisors.}
\label{fig:9}
\end{figure}

The results of the correlational analysis among advisor-advisee's collaboration duration, advisors' academic ages, proportion of advisees' h-index ranking the top 10\% are shown in Fig. \ref{fig:9}. The data in first three years of collaboration with advisors whose academic age is 5 are missing because of the slow growth rate in this period. It is difficult to calculate the top 10\% advisees' h-indices because the values of their h-indices are relatively low and similar. For $Res$ advisors, the proportion of advisees ranking top 10\% is shown in the left subgraph in Fig. \ref{fig:9-a}, and the result of $Top_{10}$ advisors is summarized in the right subgraph. Fig. \ref{fig:9-b} shows the proportion of differences between advisees mentored by $Top_{10}$ advisors and $Res$ advisors. The values of advisors whose academic ages are 30 in Fig. \ref{fig:9-b} are always below 0.1. For others, their values are all above 0.1, even the maximum value can reach 0.28. It indicates that most $Top_{10}$ advisees are coached by $Top_{10}$ advisors.

Selective bias is a possible explanation for these phenomena. The advisor-advisee relationship is a two-way choice relationship, quality advisees tend to seek top advisors and vice versa. Moreover, we have also found that the gap between these advisees is closely related to the advisors' academic age and the time duration of their collaboration. In the first three years of collaboration, the gap between advisees is relatively small. Since the fourth year, the gap begins to grow wider.

\subsection{Publication rates analysis}
As mentioned before, we have separated advisors according to their $AA$ in the experiment. However, considering the publication rates increased over time, it may have a biased effect on the scholars' number of publications, citations, and h-indices. So we carry out the experiment to verify whether the publications rates impact our findings.

\begin{figure}[htbp]
\centering
\subfigure[New records per year ]{
\label{fig:12-a}
\includegraphics[width=0.48\textwidth]{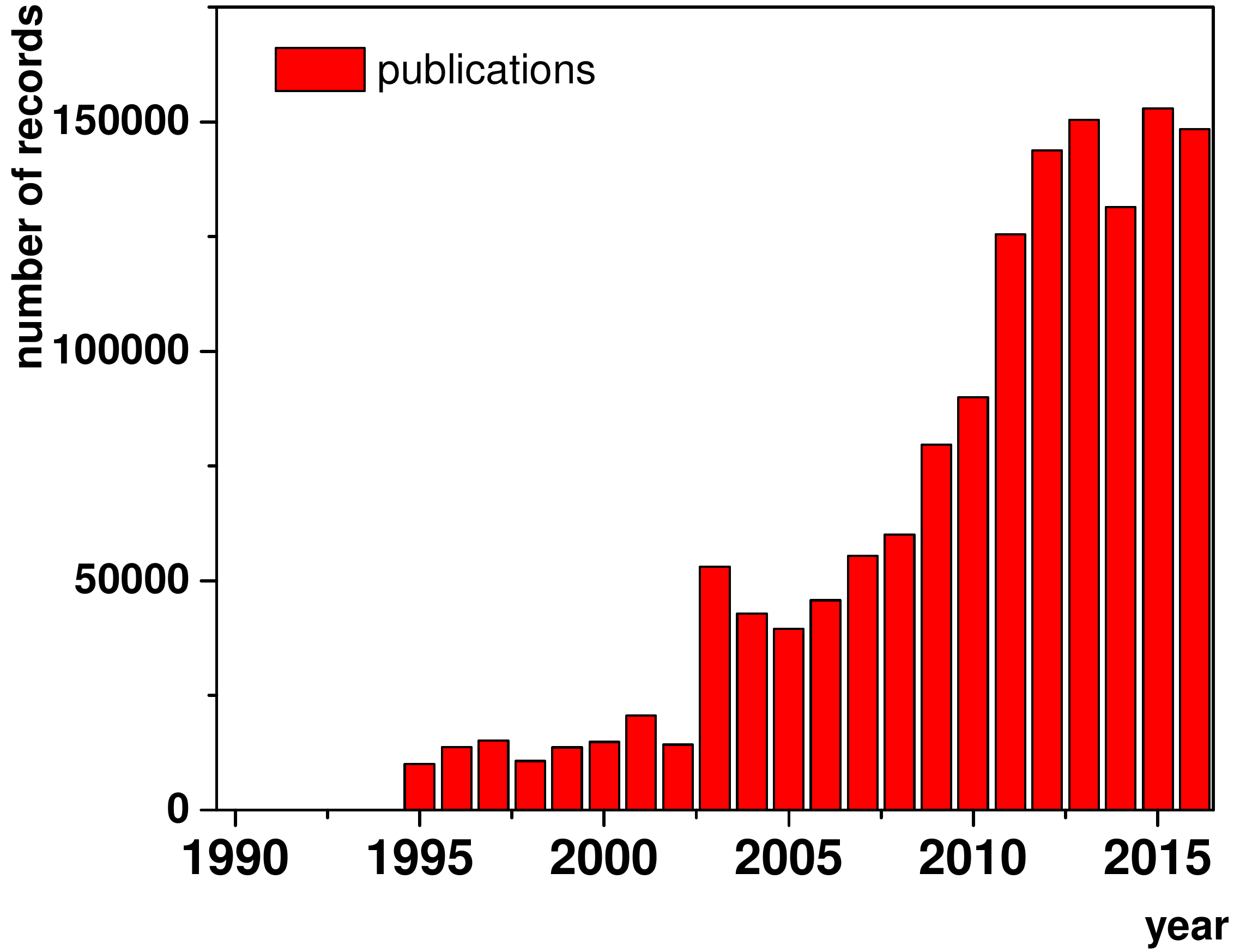}}
\subfigure[1996-2002]{
\label{fig:12-b}
\includegraphics[width=0.46\textwidth]{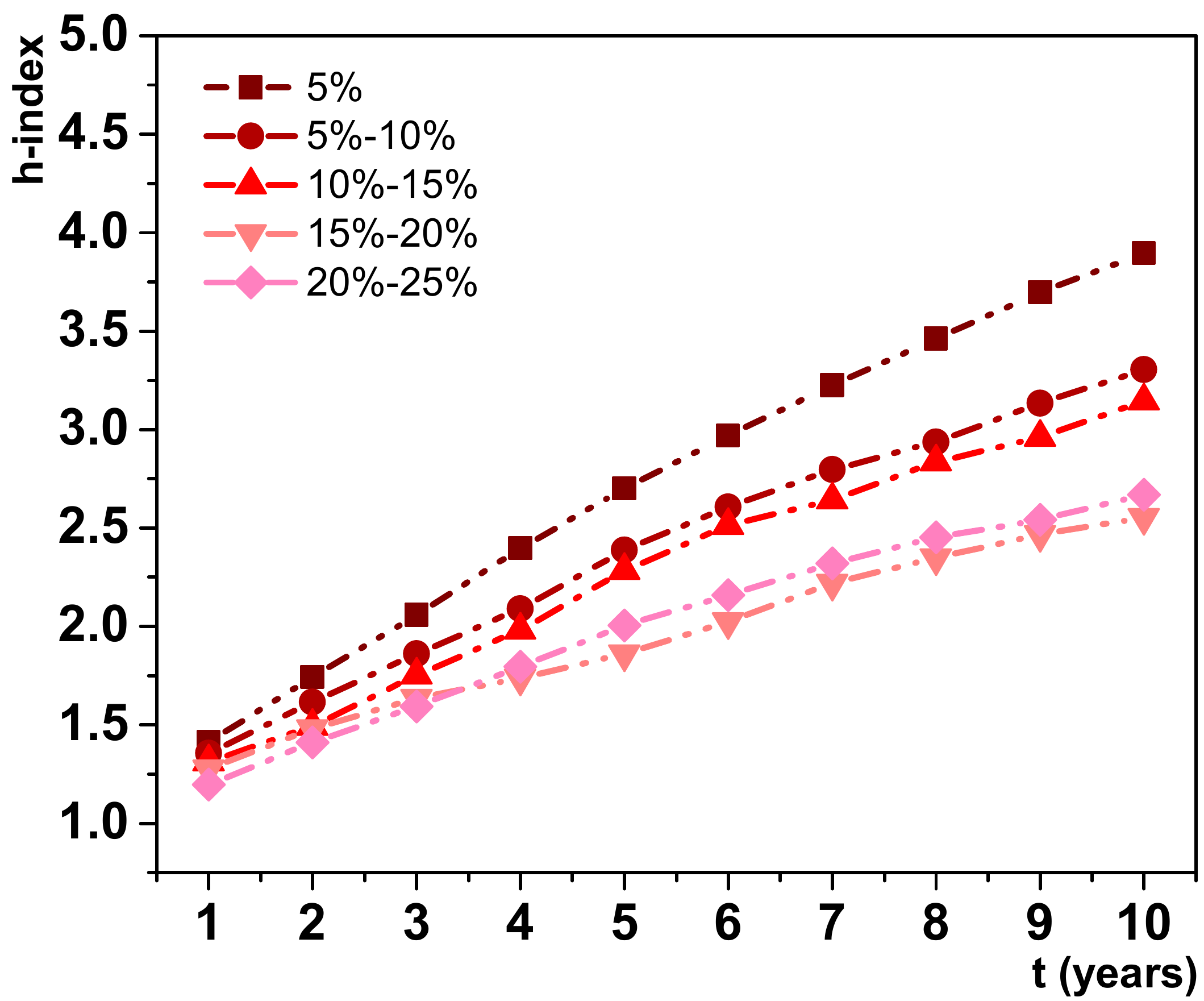}}\\
\subfigure[2003-2007]{
\label{fig:12-c}
\includegraphics[width=0.46\textwidth]{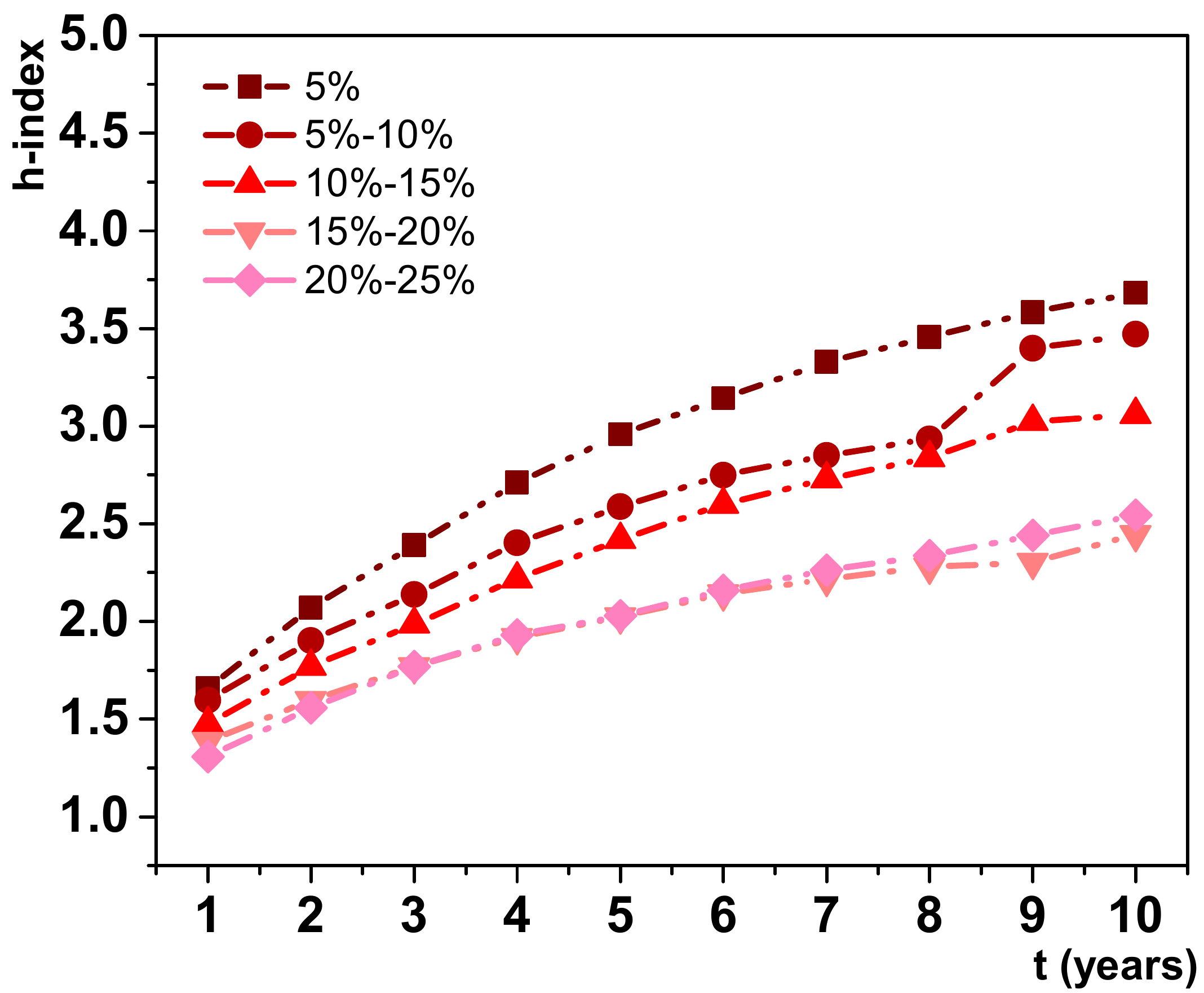}}
\caption{Advisees' h-indices whose advisors' $AA$ is 20 in different time periods, as well as their h-indices are ranking at the top 5\%, 5\%-10\%, 10\%-15\%, 15\%-20\%, 20\%-25\%. (b) Advisees' average h-index mentored by advisors whose $AA$ is 20 in 1996-2002. (c) Advisees' average h-index mentored by advisors whose $AA$ is 20 in 2003-2007.}
\label{fig:12}
\end{figure}

Fig. \ref{fig:12-a} displays the number of new entries in the database per year. From the data in Fig. \ref{fig:12-a}, it is apparent that the growth rate of the publication is different at different stages. The number of new records increases sharply in 2003 and 2011. To be more specific, there are less than 25,000 new publications per year in 1999-2002 but over 50,000 in 2003-2007, almost two times more than the first period. In order to ensure all of the advisees have plenty of time to accumulate enough publications, we choose advisors whose academic age is 20 in 1996-2002 and 2003-2007, respectively. We compare the average academic performance (h-index) of their advisees. In Fig.~\ref{fig:12-b} and  Fig.~\ref{fig:12-c}, the polylines show the advisees' average h-index mentored by different h-index ranking advisors. Data in these figures can be compared with the data in Fig. \ref{fig:6}, which is the basis of our findings. A positive correlation can be found between advisors' and advisees' academic performance. So the publication rates have limited influence on the phenomena we have found in this work.

\section{Discussion}

Moreover, we have processed our dataset and do experiments on the part of the real dataset to explore the correlation between the advisor and their academic grandchildren. We extract 4,256 advisor-advisee pairs (645 advisors) and compile the statistics on the data. For advisors, we still divide them into two groups according to their h-indices: $Top_{10}$ advisors and $Res$ advisors. And then we analyze their grandchildren's impact. We discover that the average h-index of $Top_{10}$ advisors' grand-advisees is 40\% higher than the grand-advisees mentored by Res advisors. It illustrates that there may be also a correlation existing between them and it needs in-depth research. Usually, PhD students are guided directly by their advisors, thus we only focus on the direct advisor-advisee relationships. Through this analysis, a comparatively deep analysis should also be made on the relationship between the advisor and their academic grandchild. We will carry out an in-depth study of this issue in future work.

There are still a few limitations in this work. Firstly, the conclusions are efficacious only for Computer Science, which is a rapidly developing discipline and has the characteristic of spreading its knowledge in conferences. It would be interesting to explore the anatomy of the relationship in other traditional areas of knowledge if we can get the large-scale advisor-advisee relationships dataset for the discipline. Secondly, we only show the correlation between advisors' academic characteristics and advisees' academic performance. Other factors may also produce these phenomena, such as selective bias and the rank of institutions. It is necessary to further explore the cause of these phenomena. Thirdly, it is unfortunate that the study only illustrates the advisors' benefits derived from advisees on the formal relations. In terms of relevant experience, the informal advisor-advisee relationship is also an important part of mentorship. In future work, through the continuous improvement of the dataset, we will conduct the in-depth study on the informal advisor-advisee relationships.

\section{Conclusion}
This study presents advisees' potential academic performance of choosing different advisors with different academic ages or academic levels in the field of Computer Science. It has identified the relationship between
advisors' academic age and advisees' performance, which experiences an initial growth, follows a sustaining stage, and finally ends up with a declining trend. The second major finding is that accomplished advisors can bring up skilled advisees, which is evidenced by advisees' academic performance advised by different advisors. Taken together, these findings suggest a significant role for advisors in promoting their advisees. This research extends our knowledge of scholars' career success and will serve as a base for future studies.

In summary, through the analysis of the relationship between advisors and advisees, the findings have significant implications to understand the relationship between advisors' academic characteristics and advisees' performance. Moreover, it can shed light on the advisor recommendation. However, it should be noted that the results may differ in different research fields, it would be interesting to explore the anatomy of the relationships in other areas of knowledge.

\section*{Acknowledgment}
The authors extend their appreciation to the Deanship of Scientific Research at King Saud University for funding this work through research group NO (RGP-1438-27).

%



\end{document}